\documentclass{ws-rv961x669}
\usepackage{ws-rv-van}     
\usepackage{ws-rv-thm}     
\usepackage{subfigure}     
\makeindex

\newcommand{\araa}{Annu. Rev. Astron. Astrophys.}   
\newcommand{\aj}{Astron. J.}   
\newcommand{\apj}{Astrophys. J.}   
\newcommand{\apjl}{Astrophys. J. Lett.}   
\newcommand{\apjs}{Astrophys. J. Suppl. Ser.}   
\newcommand{\ao}{Appl. Opt.}   
\newcommand{\aap}{Astron. Astrophys.}   
\newcommand{\aapr}{Astron. Astrophys. Rev.}   
\newcommand{\mnras}{Mon. Not. R. Astron. Soc.}   
\newcommand{\nat}{Nature} 
\newcommand{\nar}{New Astron. Rev.}   
\newcommand{\pasp}{Publ. Astron. Soc. Pac.}   
\newcommand{\ssr}{Space Sci. Rev.}   

\begin{document}

\chapter[]{Observations of Stellar-Mass Black Holes in the Galaxy}

\author[M. MacLeod and J. Grindlay]{Morgan MacLeod and Jonathan Grindlay }

\address{Center for Astrophysics, Harvard \& Smithsonian, \\ Cambridge, MA, 02138}

\begin{abstract}
Stellar-mass black holes (BHs), with masses comparable to stars, are a major constituent of our Milky Way galaxy. This chapter describes the landscape of challenging, and long-sought efforts to identify these objects in the Galaxy. The first stellar-mass BHs were identified as persistent, but highly variable cosmic X-ray sources. Later, transient BH candidates were detected, and now far outnumber the persistent sources. Decades of effort have also yielded candidate BHs via gravitational microlensing and their orbital effect on binary companions. Populations of BH systems have begun to emerge from these detection strategies, offering insight into the astrophysical context in which BHs exist and driving questions about the formation, assembly, and ongoing evolution of these enigmatic objects.
\end{abstract}


\body

\section{Introduction} \label{sec:intro}

Black holes (BHs) are thought to be natural remnants of the stellar evolution of massive stars (see Chapter 3 in this volume). Given the history of star formation in the Milky Way, there are likely to be up to $10^8$ such BHs in our Galaxy.\cite{1983bhwd.book.....S} And yet, without a power source from luminous accretion, these sources are intrinsically dark, and difficult to detect.\cite{2006csxs.book..507K} This chapter describes 50 years of progress discovering these sources in the Galaxy through their X-ray and optical accretion luminosity and, more recently, through their gravitational effects on binary companions and background stars. And yet, at present, we have only dynamically confirmed a few tens of BHs in the Galaxy! If the predicted population of BHs is accurate, the discovered sources represent a tiny fraction of the entire population of BHs.

The first stellar mass BHs were discovered as luminous X-ray sources produced by accretion of matter on to a BH in orbit around a more massive star.\cite{2006csxs.book..507K,2010csxs.book..157M} X-rays are produced when the mass-rich wind of the massive star is gravitationally captured into an accretion disk around the BH. Later, sources were discovered where a lower-mass star directly overflows its Roche lobe into an accretion disk about the BH. In each of these cases, emission from the disk and its corona combine to produce broad-band X-ray emission.

X-ray quiescent BHs have now also been observed in wider, non-interacting binaries. These systems have characteristically longer orbital periods, and the BH's presence is inferred based on a mass estimate and the lack of corresponding luminous emission. Finally, a strong candidate for an isolated BH has recently been discovered by its strong gravity, which, from Einstein's general relativity, bends space around it. This gravitational microlensing brightens a normal star that happens to be behind the BH as viewed from Earth. 

The remainder of this chapter is organized as follows. In Section \ref{sec:hmxb}, we introduce the BH High Mass X-ray Binaries, which are systems containing a massive stellar companion. In Section \ref{sec:lmxb}, we discuss their lower companion-mass counterparts, where the BH directly accretes from a roughly solar-mass companion. These sources have rapidly become the most abundant known BHs and candidate BHs in the Galaxy, allowing us to consider their spatial distribution in the galaxy and differing models for their formation. In Section \ref{sec:spin}, we discuss what is known about the spins of X-ray emitting BHs in the Galaxy. In Section \ref{sec:nonxray}, we discuss the discovery of BHs in binaries through their orbital motion and as microlensing sources that transiently magnify background stars. In Section \ref{sec:conclusion}, we conclude.  

\section{Black Hole High-Mass X-ray Binaries - BH-HMXBs}\label{sec:hmxb}
\subsection {The first stellar mass black hole: Cygnus X-1}

\subsubsection{Discovery}
Cyg X-1 was discovered in X-rays in 1964 by the Naval Research Lab with modified Geiger counter X-ray detectors on a sub-orbital sounding rocket. The rocket spent only $\sim5$~min above the atmosphere, which is required to detect cosmic X-rays, but identified eight cosmic X-ray sources as it rotated to scan the sky, including Cyg X-1.\cite{1965Sci...147..394B} Cyg X-1 was then studied intensely with the first proportional counter X-ray detectors on a satellite, the UHURU mission (1970 - 1973).\cite{1971ApJ...165L..27G} UHURU discovered over 300 X-ray sources\cite{1978ApJS...38..357F}, including the first bright ``X-ray binaries" that are now thought to be neutron stars (NSs) orbiting a more massive star. Massive stars, $\gtrsim10 M_\odot$,  have strong outflowing winds driven from their surface by radiation pressure. A fraction of this mass loss is gravitationally captured into an accretion disk by a nearby compact object such as a NS.\cite{1983bhwd.book.....S} The accreting material flows toward and eventually impacts the surface of the NS, thermalizing and subsequently radiating an enormous X-ray flux.\cite{1983bhwd.book.....S,2010csxs.book..157M}

Cyg X-1 was different. It didn't pulse due to the accretion disk being funneled onto a North or South pole of the strong magnetic field that accreting NSs often have.\cite{1983bhwd.book.....S,2010csxs.book..157M} Cyg X-1 didn't eclipse as a companion star will cause if the orbital plane is viewed nearly edge-on.  Cyg X-1 variability was sporadic; no pulses. The short-timescale (seconds) variability of Cyg X-1 is accompanied by longer (weeks--years) timescales for its two X-ray ``spectral states" to inter-change.\cite{1974ApJ...189L..13R,2014wbll.book.....B} On $\sim$ 6 - 10 year timescales, Cyg X-1 changes from a ``normal" low-flux ``hard" X-ray (high energies) power law  spectrum to a much larger flux of soft (low energies) X-ray thermal spectrum with duration $\sim$2 - 3 years before returning to the low-hard state.\cite{2006ARA&A..44...49R} A high-low flux transition for Cyg X-1 was discovered with UHURU \cite{1972ApJ...177L...5T} to coincide with the onset of strong non-thermal radio emission. However, a  recent model\cite{2018ApJ...860..166T} of  Cyg X-1's spectral state transitions does not include the correlation with radio emission. Cyg X-1 was clearly very different from the identified NS accretion-powered pulsars. Although this theoretical picture emerged over time, from its X-ray properties alone it was not obvious that Cyg X-1 was an accretion powered X-ray binary, or a BH.

\subsubsection{Dynamical Confirmation}
A major development came when an optical counterpart of the Cyg X-1 X-ray source was identified\cite{1971Natur.233..110M}. The optical counterpart is a very luminous spectral type B0Ib supergiant (now classified as O9Iab\cite{2011ApJS..193...24S}) in the Cygnus OB3 association. 

With optical monitoring, the techniques used to study binary stars and their masses may be employed to constrain the masses of putative BHs in binary systems. Both photometric and radial velocity observations have potential discriminating power in determining a system's orbital period. However, radial velocities directly probe the system's dynamics. Together the orbital period, $P$, and the semi-amplitude of the system radial velocity curve, $K$, determine the binary's mass function
\begin{equation}
    f = \frac{P K^3}{2\pi G} = \frac{M_x \sin^3 i}{(1+q)^2}
\end{equation}
where $i$ is the orbital inclination angle of the binary ($i=90$ degrees is viewed edge on, maximizing the radial velocities), $M_x$ is the mass of the X-ray source that is unobserved in the optical, and $q=M_\ast/M_x$, the star to compact object mass ratio.\cite{2014wbll.book.....B} Several parameters here are typically not known a priori: these are the inclination and the mass ratio. 

Despite uncertain mass ratio and inclination, it is always true that $f\leq M_x$. Thus, the measurement of a mass function alone via the radial velocity semi amplitude and period can establish a dynamical mass lower limit on the putative BH in an X-ray binary. Remillard and McClintock\cite{2006ARA&A..44...49R} tabulate the mass functions of known BH systems, many of which exceed $3M_\odot$. Measuring the component masses individually requires constraints on inclination and mass ratio.

Several methods have been employed to constrain inclinations and mass ratios in binary systems. One is stellar rotation. Under the assumption that a donor star is tidally locked, its rotation period matches the orbital period. It should have an equatorial velocity of $V_{\rm rot} = 2\pi R_\ast/P$, where $R_\ast$ is the stellar radius. The same spectroscopic observations that are used to probe radial velocities can potentially assess $V_{\rm rot}$ through  the broadening that is observed averaging over the stellar disk.\cite{2014wbll.book.....B} In cases where a distance is known, and thus $R_\ast$ can be estimated from the stellar surface temperature and apparent brightness, this known rotational Doppler broadening of spectral lines places a powerful constraint on the inclination. 

Additionally, many X-ray binary systems are so close that the stellar component is distorted by tides from its binary companion. When viewed with $i>0\deg$, this distortion of the star projects varying surface area toward the observer as a function of orbital phase. At a given stellar radius, the tidal vulnerability of the companion star is proportional to the mass ratio $q$. Cases where radial velocities, rotational broadening, and ellipsoidal variability are all observed offer a variety of independent constraints by which a BH mass may be constrained. Presently, software is usually used to perform a joint fitting of the binary properties under the known constraints.\cite{2018maeb.book.....P}

Returning to the case of Cyg X-1, an orbital period of 5.6 days is seen both in ellipsoidal photometric variability of the companion star, which is distorted by the BH's tidal field and in radial velocities, which probe the projection of the companion star's orbital motion onto our line of sight\cite{2021Sci...371.1046M}. The photometric variability has a semi-amplitude of approximately 0.02~mag in the optical while the radial velocity semi-amplitude is $\sim 75$~km~s$^{-1}$ (Miller-Jones et al. provide a recent summary of currently available data on the Cyg X-1 binary\cite{2021Sci...371.1046M}). 

\subsubsection{Eventual Model and Recent Mass Revision}
Over the 50 years since its discovery, a cohesive model for Cyg X-1 has emerged. The Cyg X-1 X-ray emission is thought to be due to accretion of the strong stellar wind from the massive O9Iab super-giant  star on to a BH. The BH mass was initially derived by measuring the mass function of the 5.6 day orbital period binary and the orbit inclination\cite{2011ApJ...742...84O} as $14.8\pm 1 M_\odot$. This was the first definitive stellar mass BH, with mass $\gg3M_\odot$, the approximate limiting mass for a neutron star (NS) to not collapse to a BH.\cite{1983bhwd.book.....S,2014wbll.book.....B} The Cyg X-1 BH mass was revised in 2021 to be $21.2\pm2.2M_\odot$ and the donor O star to be $41\pm7M_\odot$ because the distance to the binary was derived to be larger than previous estimates\cite{2021Sci...371.1046M}. The larger distance (of 2.2~kpc instead of 1.86~kpc) is inferred from radio parallax measurements and Gaia optical parallaxes\cite{2021Sci...371.1046M}. The radio astrometry also resolves the orbital motion of the BH on the sky, and can be jointly fit with the radial velocities of the companion. Together these constraints imply a more massive system than previously believed.\cite{2011ApJ...742...84O,2021Sci...371.1046M}

\subsection{Discovery of other BH-HMXBs}

UHURU and subsequent X-ray satellites soon identified two additional Galactic BH-HMXBs with properties very similar to Cyg X-1. These were LMC X-1 and LMC X-3, both located in the Large Magellanic Cloud (LMC).\cite{2009ApJ...697..573O,2014ApJ...794..154O} As young, massive systems, BH-HMXBs  are found in regions of ongoing star formation, OB associations. They necessarily have lifetimes limited by the lifetime of their massive-star companions to a few million years. The LMC is a host to ongoing star formation at a rate\cite{2009AJ....138.1243H} of $\sim 0.2 M_\odot$~yr$^{-1}$. When multiplied by an estimated BH-HMXB age of $5\times 10^6$~yr, we see that the two known LMC BH-HMXBs represent a fraction of $\sim 10^{-4}$ to $10^{-5}$ of the $10^6M_\odot$ of recent star formation. This fraction can be compared to the mass fraction of roughly $10^{-1}$ that goes into stars $>20M_\odot$ in a Salpeter initial mass function (for which $dN/dm \propto m^{-2.3}$).  Thus, empirically, one in every 1,000 to 10,000 massive stars formed is found in a BH-HMXB state.

 There are other massive stars in our MW that are in X-ray binaries that have evaded definitive dynamical confirmation. These systems have such strong mass loss, due to their extreme luminosity, that velocity measurements of the companion stars have proven challenging. For example, Cyg X-3 is almost certainly a BH-HMXB but its BH is ``fed" by a super-luminous wind from its Wolf-Rayet star binary companion.\cite{2013MNRAS.429L.104Z} 

 The most extreme candidate BH-HMXB in the MW is SS433, the 13.1 day binary that launches mass from its supergiant companion star wind into a nearly relativistic jet just before accreting on to the presumed BH companion.\cite{1984ARA&A..22..507M,2018MNRAS.479.4844C}  The jet is launched perpendicular to the orbital plane of the massive supergiant A star and BH binary with  velocity $\sim0.26$ of the speed of light. The binary accretion disk and central jet precess with a period of 162.5d. This first ``microquasar" has finally been understood by extensive optical spectroscopy of the  changing profiles of multiple emission lines from the disk (HI, HeI, OI, SiII, CaII and FeII) and absorption lines from the A-star companion. This spectroscopy allowed the mass of the BH, $4.2\pm 0.4M_\odot$, and the supergiant A star, 11.3 $\pm 0.6 M_\odot$, to be derived\cite{2020A&A...640A..96P} However, these determinations are not universally accepted. For example, in a recent review Cherepashchuk\cite{2021Univ....8...13C} argues for BH mass of $>8M_\odot$ and that the mass-rich wind complicates the interpertation of spectral data.

\begin{figure}
     \centering
     \includegraphics[width=0.7\textwidth]{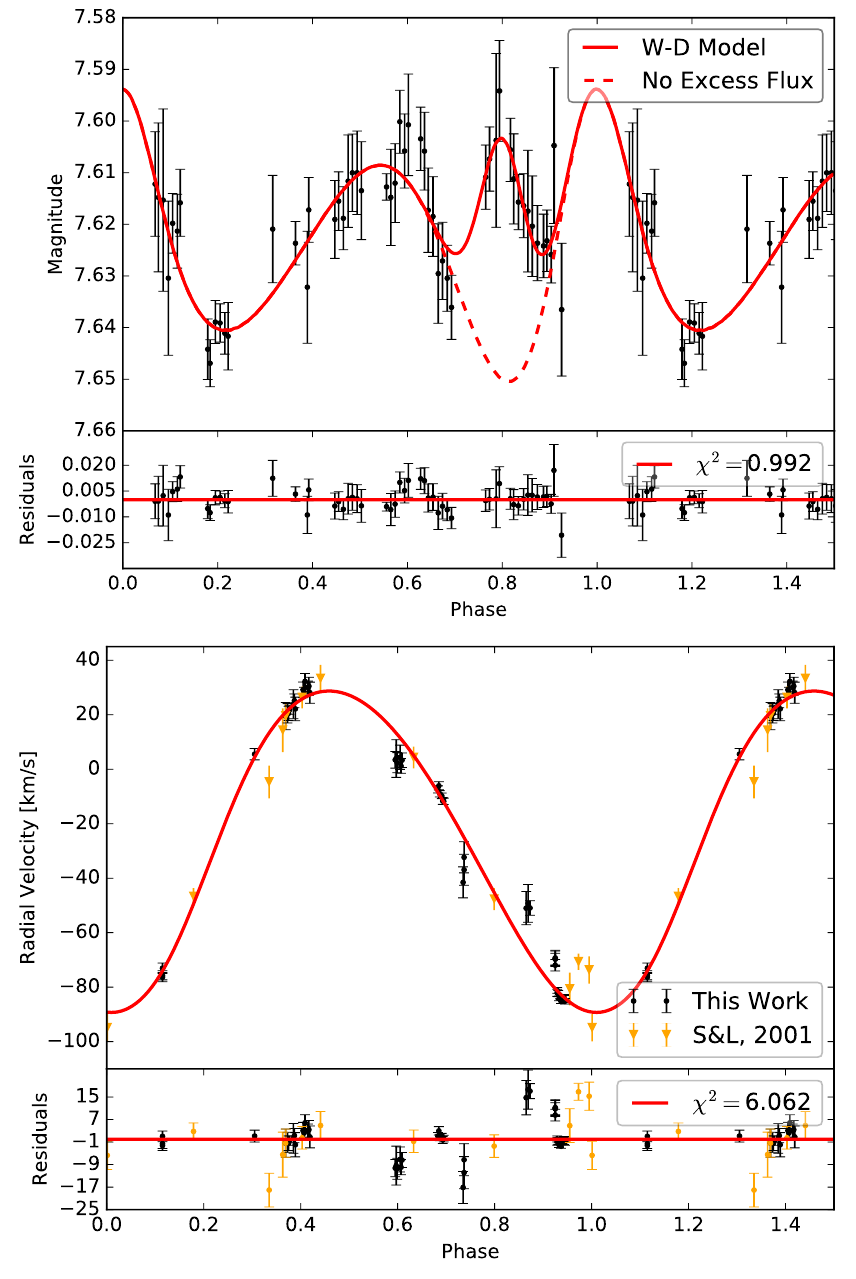}
     \caption{Light and radial velocity curves of the HD96670 single-line spectroscopic binary.  This figure illustrates an example of how ellipsoidal variations and radially velocities, when jointly fit, can constrain system parameters like the mass ratio, inclination, and component masses. The feature at phase of 0.8 is excess optical flux from the reduced Kramers opacity of the wind by X-ray heating from the bow shock and accretion disk at superior Conjunction. Figure reproduced with permission from  Gomez and Grindlay\cite{2021ApJ...913...48G}. }
     \label{fig:gg}
 \end{figure}

 Recently, Gomez and Grindlay reported on optical monitoring of HD96670, a single-line spectroscopic binary in the Carina OB2 association.\cite{2021ApJ...913...48G}. From joint modeling of the light and radial velocity curves, a period of $5.28$~d and component masses  of $22.7^{+5.2}_{-3.6}M_\odot$ and $6.2^{+0.9}_{-0.7}M_\odot$ were identified\cite{2021ApJ...913...48G}. The $\sim 22M_\odot$ object is a luminous ``weak wind" O8IV star, which produces the spectroscopic lines observed. The $\sim 6M_\odot$ companion is unobserved in optical, and is a BH since the binary was first observed with NuSTAR as a possible BH-HMXB and found to have a power law spectrum from 3 - 20 keV which rules out an optical star companion but is consistent with a BH. Further X-ray observations are planned to test the accretion model and optical ``flare" at RV phase 0.8 as being due to reduced Kramers opacity  by X-ray heating of  the stellar wind at superior conjunction. Figure \ref{fig:gg} shows the system's phased radial velocity and light curves along with the joint dynamical fitting.\cite{2021ApJ...913...48G}

 Weak wind O stars  (O8-O9 main sequence or sub-giant luminosity classes)  should be $\gtrsim30\times$ more numerous than super-giants due to their respective lifetimes. BH systems containing these stars have not been discovered with wide-field X-ray missions due to their low luminosity ($L_x \sim 10^{31-32}$~ergs s$^{-1}$), but can be discovered by future wide-field X-ray imaging missions which can target nearby OB associations.  
 
\section{Black Hole Low-Mass X-ray Binaries - BH-LMXBs}\label{sec:lmxb}
 
\subsection{Discovery of A0620-00 and Black Hole Transients}

\begin{figure}[tbp]
    \centering
    \includegraphics[width=\textwidth]{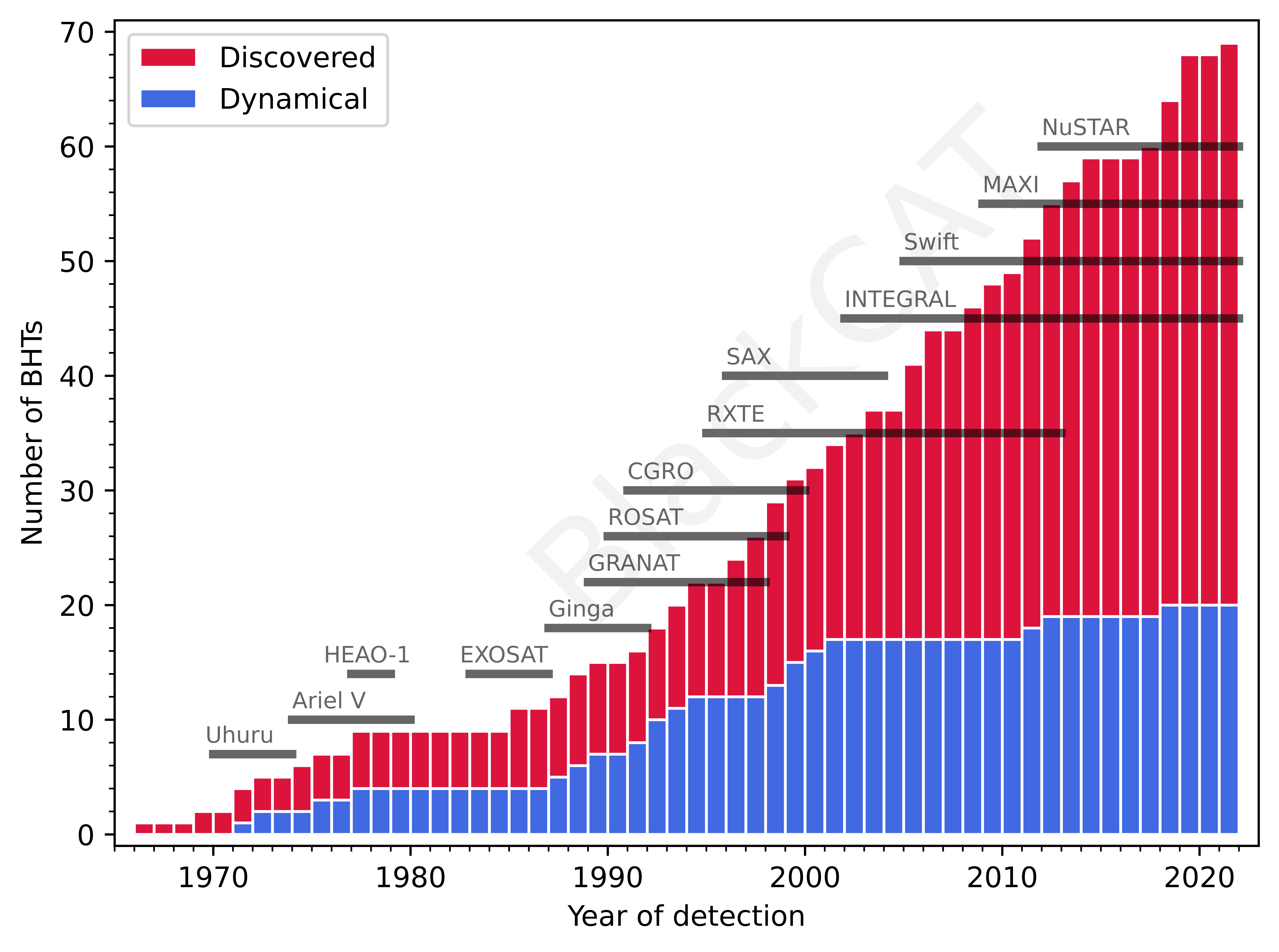}
    \caption{Cumulative number of BH candidates (red) and vs. those with dynamically confirmed BH counterparts (blue) as a function of year of discovery. Figure adapted with permission from Figure 1 of Corral-Santana et al\cite{2016A&A...587A..61C}. The X-ray missions with large sky coverage and/or followup capability are shown across the top with horizontal black lines showing their mission duration. Systems labeled discovered in red are transients thought to be associated with outbursts of accreting BHs. Systems marked dynamical are those binaries that have been dynamically confirmed via follow up observations. }
    \label{fig:cumulative}
\end{figure}

The discovery\cite{1975Natur.257..656E}  of the extremely bright X-ray Transient, A0620-00, eventually heralded the discovery of the first BH-LMXB -- though the system was not recognized as such until an optical counterpart and dynamical confirmation followed. 
The mass function of the binary and minimum mass for the compact object was measured to be $>3M_\odot$ and thus a BH.\cite{1986ApJ...308..110M} Its current best determination is $6.6\pm0.3 M_\odot$\cite{2016ApJS..222...15T}. 
A0620-00 was different than Cyg X-1 and LMC X-3, which were confirmed earlier in that its X-rays were transient. Two outbursts are known to have occured, the first is identified in the optical on Harvard photographic glass plates\cite{1976ApJ...203L..17E} in 1917, while the second was first identified in X rays in 1975\cite{1975Natur.257..656E}.

To date, 68 candidate BH transients have been identified, as tabulated in the continuously-updated BlackCAT catalog\cite{2016A&A...587A..61C}. A similar catalog was compiled in 2016 by Tetarenko et al\cite{2016ApJS..222...15T}. The cumulative number of these discoveries is shown in Figure \ref{fig:cumulative}. Figure \ref{fig:cumulative} highlights that many more BH transients have been discovered than have been eventually dynamically confirmed. BH transients and their corresponding candidate BHs are thus the most common BH candidates known in the Galaxy. 

\subsection{Disk Instability Model and Outbursts}

\begin{figure}[tbp]
    \centering
    \includegraphics[width=0.6\textwidth]{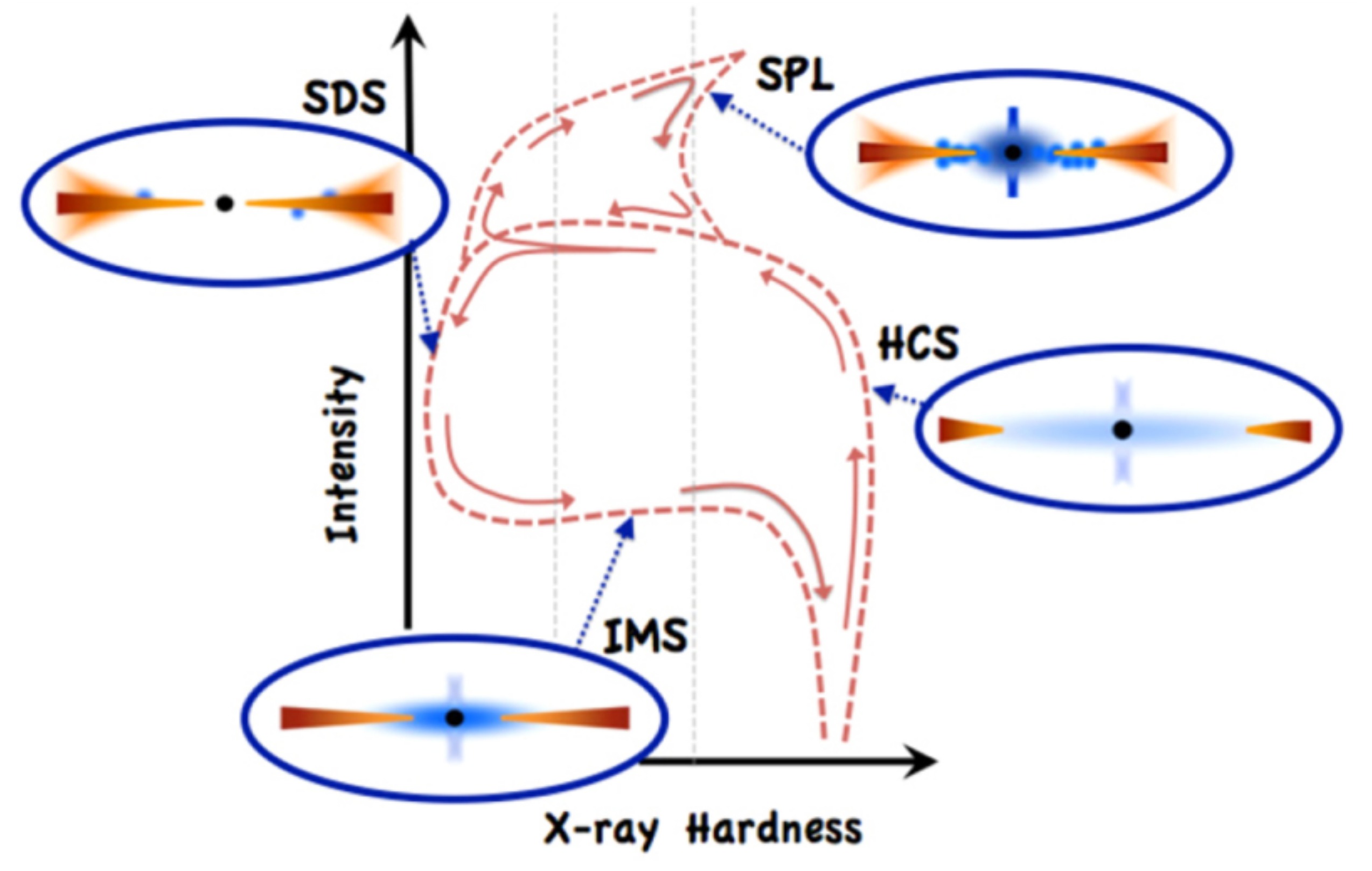}
    \caption{A cartoon rendering of changes in X-ray luminosity and spectral hardness as a BH transient undergoes outburst. Several states are labeled including the soft disk state (SDS), intermediate state (IMS), hard Comptonized state (HCS) and steep power law (SPL). The distinction in accretion state accompanies changes in accretion rate as the binary undergoes outburst.  Figure is reproduced with permission from Tetarenko et al\cite{2016ApJS..222...15T}.  }
    \label{fig:outburstcartoon}
\end{figure}

A cartoon of a BH transient outburst in the X-ray is shown in Figure \ref{fig:outburstcartoon} adapted from Tetarenko et al\cite{2016ApJS..222...15T}. Systems undergo state transitions as the accretion rate and X-ray intensity change. In these cases, several different X-ray morphologies have been identified and categorized\cite{2006ARA&A..44...49R,2010csxs.book..157M,2016ApJS..222...15T}, relating to changes in the disk and emitting region geometry as the accretion rate changes. 

\begin{figure}[tbp]
    \centering
    \includegraphics[width=0.8\textwidth]{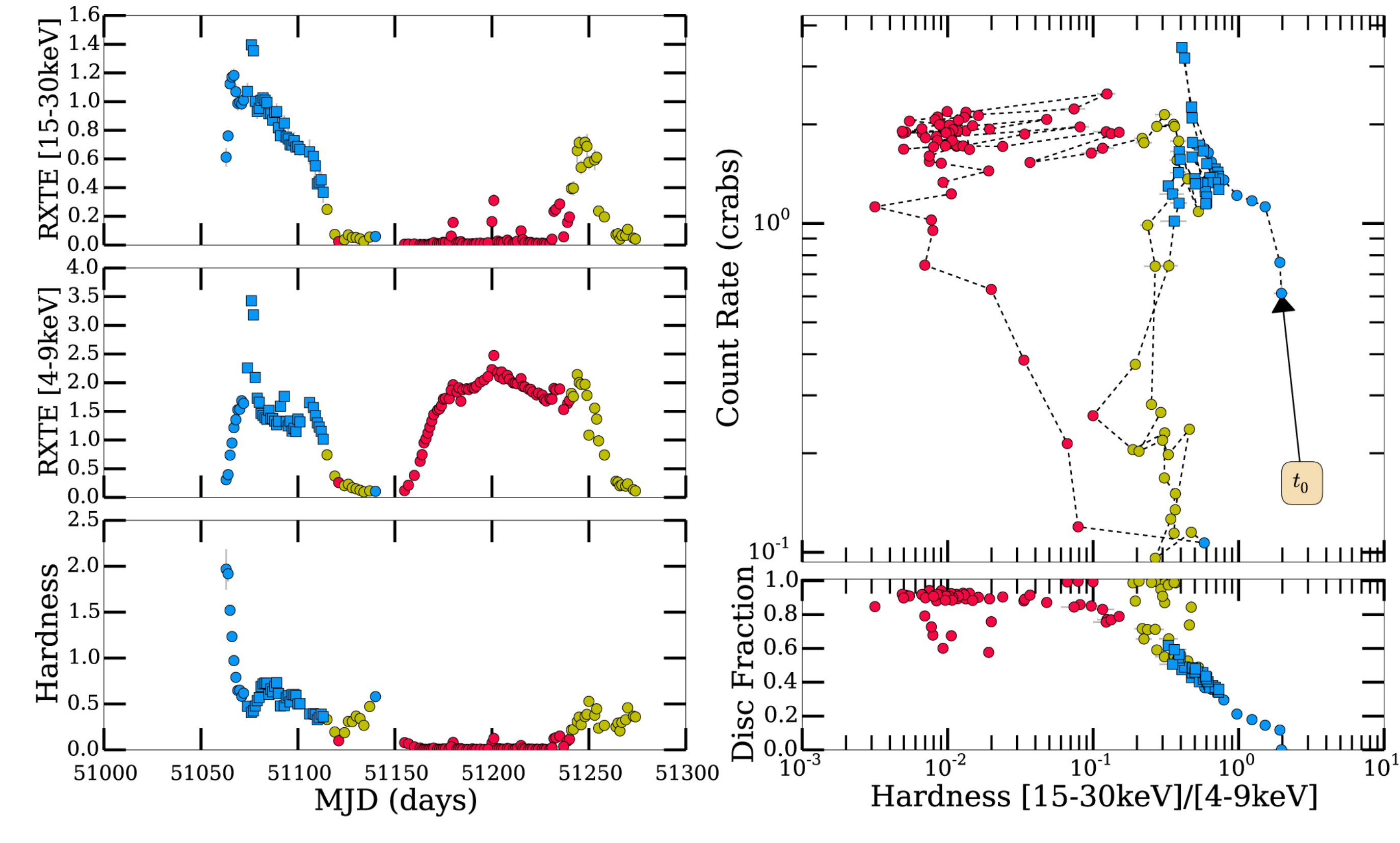}
    \caption{The 1998/1999 outburst of XTE J1550-564. Colors represent accretion state as outlined in Figure \ref{fig:outburstcartoon}, with HCS (blue), SDS (red), and IMS (yellow). As the outburst proceeds, the source transitions from a HCS upwards in intensity, through an IMS to a SDS, and eventually back down to an IMS. By probing a variety of accretion rates (which are thought to roughly track the intensity) BH transient outbursts probe accretion physics under a wide variety of conditions in a single object.  Figure is reproduced with permission from Tetarenko et al\cite{2016ApJS..222...15T}.  }
    \label{fig:outburst}
\end{figure}

 The theoretical picture that has emerged is that outbursts are produced by thermal instability in the accretion disk. In between outbursts, the accretion rate onto the BH is presumed to be very low, such that the mass in the disk grows, eventually exceeding a critical surface density and flowing rapidly onto the BH.\cite{2001NewAR..45..449L,2006csxs.book..507K,2010csxs.book..157M,2006ARA&A..44...49R} We note that a similar thermal instability model drives smaller outbursts of accreting neutron star low mass X-ray Binaries (NS-LMXBs) and accreting white dwarfs, which are called Cataclysmic Variables (CVs) or Dwarf Novae. 
 Figure \ref{fig:outburst} shows data from a 1998 outburst of XTE J1550-564, in which we see the source transitioning by orders of magnitude in hardness as the intensity changes.

\begin{figure}[tbp]
    \centering
    \includegraphics[width=0.9\textwidth]{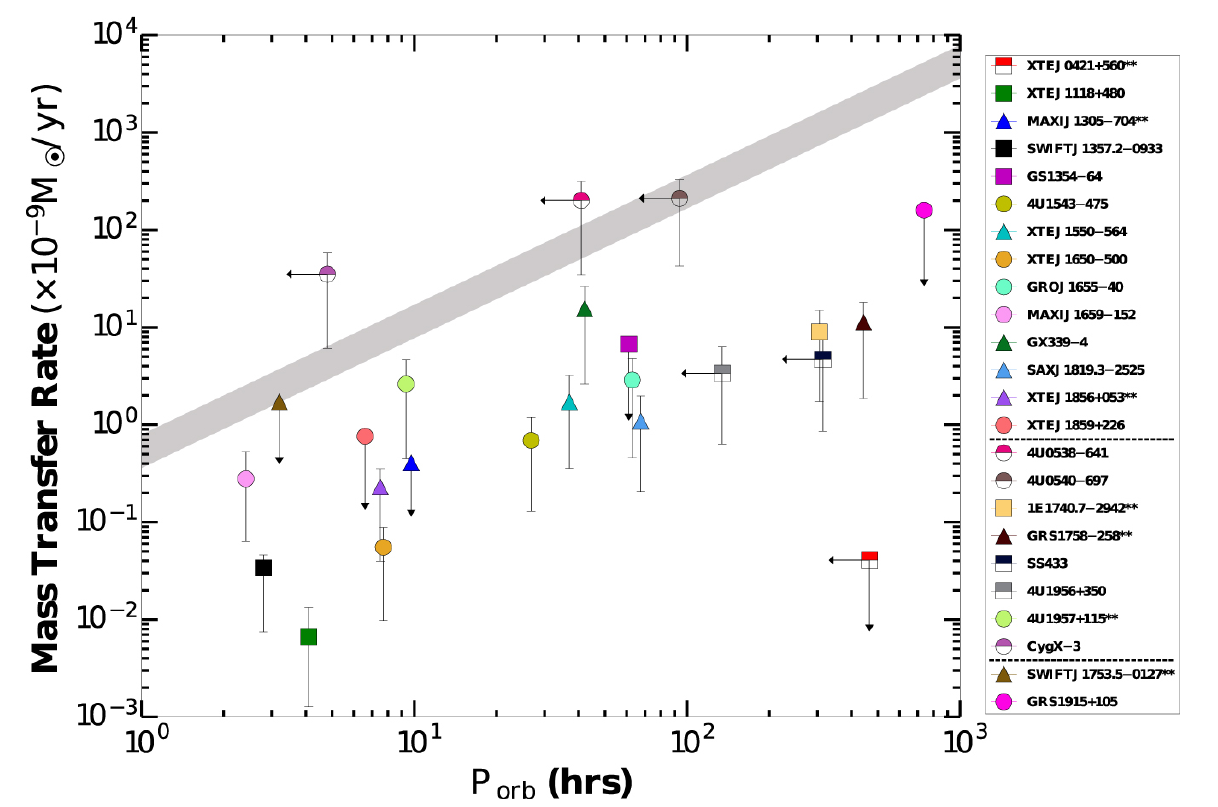}
    \caption{  Inferred mass transfer rates as a function of orbital period. BH-LMXBs are represented with fully-filled points, while BH-HMXBs have half-filled points. The grey shaded band distingishes the critical accretion rate for stable (above) or intermittent (below) accretion. That most of the BH-LMXBs accrete well below the critical value is related to their outbursting and transient behavoir. Figure is reproduced with permission from Tetarenko et al\cite{2016ApJS..222...15T}.  }
    \label{fig:mdot}
\end{figure}

\subsection{The Galactic population of BH-LMXBs}
\subsubsection{Dynamically Confirmed BH-LMXBs}

Dynamical confirmation of both BH-LMXBs (e.g. A0620-00) and BH-HMXBs (e.g. Cyg X-1) requires that the mass function of the binary be derived as described in Section \ref{sec:hmxb}. At present, the most up-to-date source of BH-LMXB candidates and dynamical confirmations is the BlackCAT catalog maintained by Corral-Santana at \url{https://www.astro.puc.cl/BlackCAT}\cite{2016A&A...587A..61C}.

There are 19 dynamically confirmed BH-LMXBs at the time of this writing. They range in mass function from 0.25$M_\odot$ to 11$M_\odot$.\cite{2016A&A...587A..61C} Of these, 15 have mass functions larger than the $3M_\odot$ value that has traditionally distinguished BHs from lower-mass compact objects\cite{2006ARA&A..44...49R}, making them unambiguous BHs even if the inclination were unconstrained. The systems have orbital periods that range from a few hours to 812~h. 

There exists remarkable uniformity in the spectral types of companions to BH-LMXBs. They range in spectral type from B9 to M, but the vast majority, 14 out of 19, are G to M dwarfs.\cite{2006ARA&A..44...49R,2016A&A...587A..61C,2016ApJS..222...15T} These stars most likely are main sequence stars that have masses of $<1M_\odot$. These systems are also the ones grouped around orbital periods less than about 10~h. At longer orbital periods of tens to hundreds of hours, the remaining five sources have higher luminosity class and are subgiants or giants.

\subsubsection{The Duty Cycle of BH-LMXB Major Outbursts}
BH-LMXBs erupt as brief transients interspersed by long periods of quiescence. Estimates of either the duty cycle of BH-LMXB outbursts or their typical recurrence time can be used to infer the Galactic population of sources\cite{2018MNRAS.474...69A}. 

The duty cycle is the fraction of time that the source is active above some threshold in a given wavelength. For a sample of observational data -- in this case, as provided by X-ray sky surveys \cite{2015ApJ...805...87Y} and by the century-long archive of photographic plates at Harvard\footnote{http://dasch.rc.fas.harvard.edu/} being digitized in the Digital Access to a Sky Century @ Harvard (DASCH) project\cite{2012IAUS..285...29G} -- the duty cycle is similar to the number of detections divided by the number of observations, 
\begin{equation}
    f_{\rm on}(>{\rm thresh}) \approx \frac{N_{\rm det} (>{\rm thresh})}{N_{\rm obs} (>{\rm thresh}) } ,
\end{equation}
where $N_{\rm det} (>{\rm thresh})$ is the number of data points in which the source is detected (above the detection flux threshold or apparent magnitude limit) and  $N_{\rm obs} (>{\rm thresh}) $ is the total number of observations with a limit that exceeds the threshold. Therefore, $N_{\rm obs} (>{\rm thresh}) $ includes any points in which the source is detected as well as any non-detections.  This expression is an approximation because a realistic set of observations contains heterogeneous detection thresholds across different observational epochs, and may contain variable observational cadence relative to the total event duration. Forward Monte Carlo modeling by injecting hypothetical events into a model of survey properties is the best way to account for these effects estimate the true value of $f_{\rm on}$, along with its uncertainty. 

The duty cycle relates the total population number of sources to the number that are active (again, above the threshold) at a given moment,
\begin{equation}
    N_{\rm on} = N_{\rm pop} f_{\rm on}.
\end{equation}
This implies that if we have an estimate of $N_{\rm on}$ and $f_{\rm on}$, we can compute $N_{\rm pop}$. To give a concrete example, if $f_{\rm on}=10^{-2}$, and $N_{\rm on}=1$, then $N_{\rm pop}=10^2$. This example implies that if a typical BH-LMXB is in outburst 1\% of the time, and at a given moment there is a 100\% chance that there is an observable BH-LMXB in outburst, then the total population of BH-LMXBs numbers 100. 

Another way to estimate the observable BH-LMXB population is to estimate about how frequently, on average, outbursts occur. If we define the recurrence time as the average time between outbursts in a given source, then the rate of outbursts (per source) is 
\begin{equation}
    \Gamma \approx t_{\rm rec}^{-1}~{\rm source}^{-1}~{\rm yr}^{-1} .
\end{equation}
The number of flares observed from a population is then
\begin{equation}
    N_{\rm flare} \approx \Gamma \Delta t_{\rm obs} f_{\rm sky} N_{\rm pop},
\end{equation}
where $\Delta t_{\rm obs}$ is the cumulative duration of observation. Imagining perfect sky coverage ($f_{\rm sky}=1$) and inverting this, we can estimate that $N_{\rm pop} \approx N_{\rm flare} / (\Gamma \Delta t_{\rm obs})$.  To give a concrete example, if the average recurrence time of flares is is 1 every 50 years, and the monitoring period has been 25 years, then $\Gamma \Delta t_{\rm obs}\approx 0.5$. For a total of $\sim$50 observed flares, the implied population is $\sim 100$ observable BHLXMBs. 

The quantities of the recurrence time and duty cycle are currently constrained by data from the X-ray observational epoch \cite{2015ApJ...805...87Y,2018MNRAS.474...69A,2022arXiv220610053B}. For example, Tetarenko et al.\cite{2016ApJS..222...15T} measure the duty cycle and recurrence time of BH-LMXB transients in the limited time span of January, 6, 1996 – May 14, 2015 for repeated BH transient outbursts. The repeating systems have recurrence times of years and duty cycles of at least 1\%. These X-ray duty cycles are biased toward higher values by the finite monitoring time.

The major current effort to examine the long-duration of photographic historical data contained in the Harvard glass plates (1886 - 1992) is the the Digital Access to a Sky Century at Harvard, DASCH, project\cite{2012IAUS..285...29G}) which provides much longer recurrence time and duration constraints in the optical than in the present X-ray data. 
These values will be better constrained when the DASCH scanning and processing is completed and includes the most sensitive $\sim$~30000 A-plates. Scanning and processing was interrupted in April 2021 by a crash of 2 DASCH servers on the Harvard Research Computing cluster, ${\it Cannon}$. 

\subsubsection{The Spatial Distribution of BHLXBs}
Through a combination of their sky positions and distance estimates, a number of studies have examined the spatial distributions of BH-LMXBs in the Galaxy. 

\begin{figure}[tbp]
    \centering
    \includegraphics[width=0.8\textwidth]{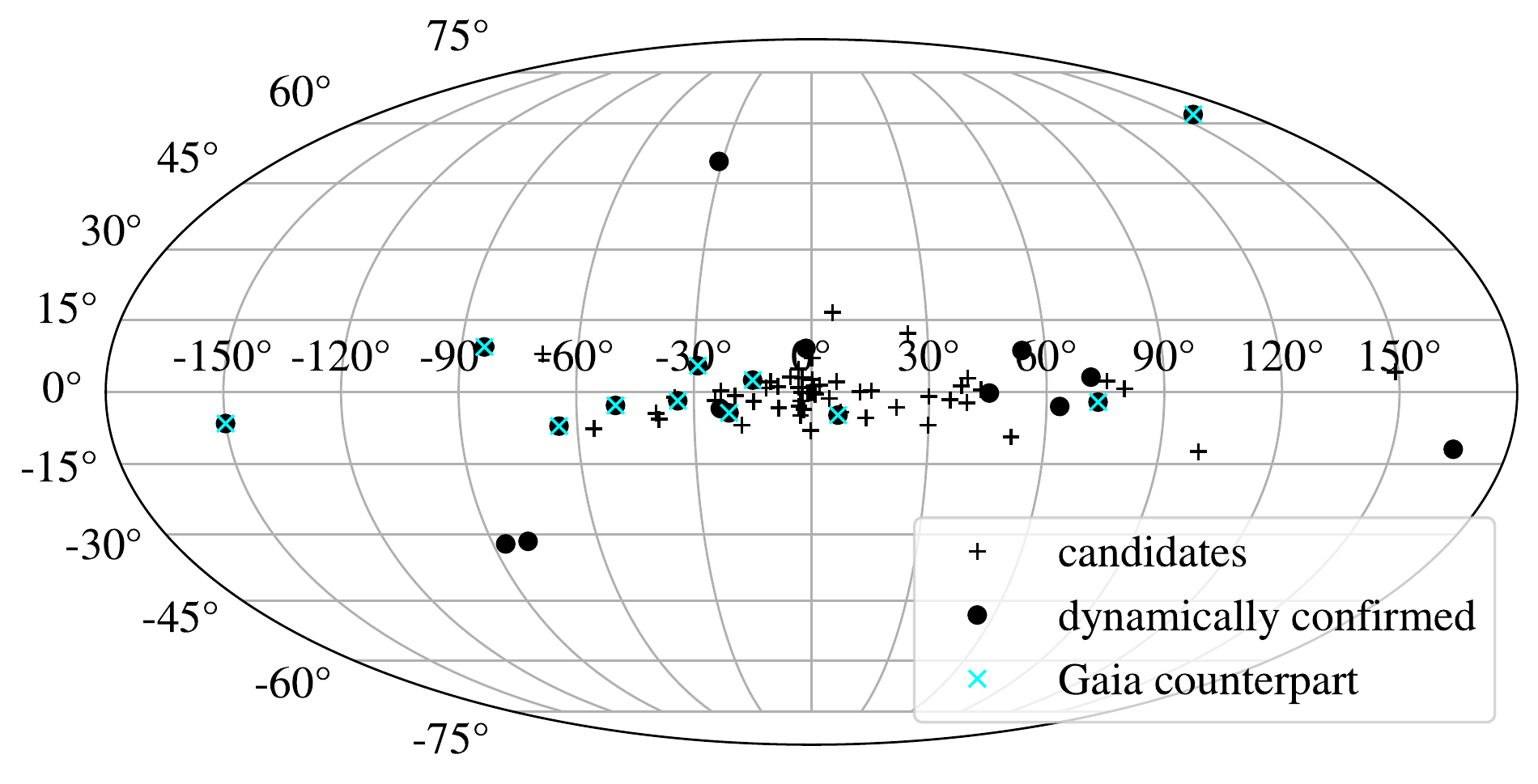}
    \caption{The sky distribution of BHs and candidates in the Watchdog catalog \cite{2016ApJS..222...15T} in Galactic coordinates. BH X-ray binaries are spread in a thick distribution around the Galactic disk and bulge.}
    \label{fig:skydist}
\end{figure}

\begin{figure}[tbp]
    \centering
    \includegraphics[width=0.6\textwidth]{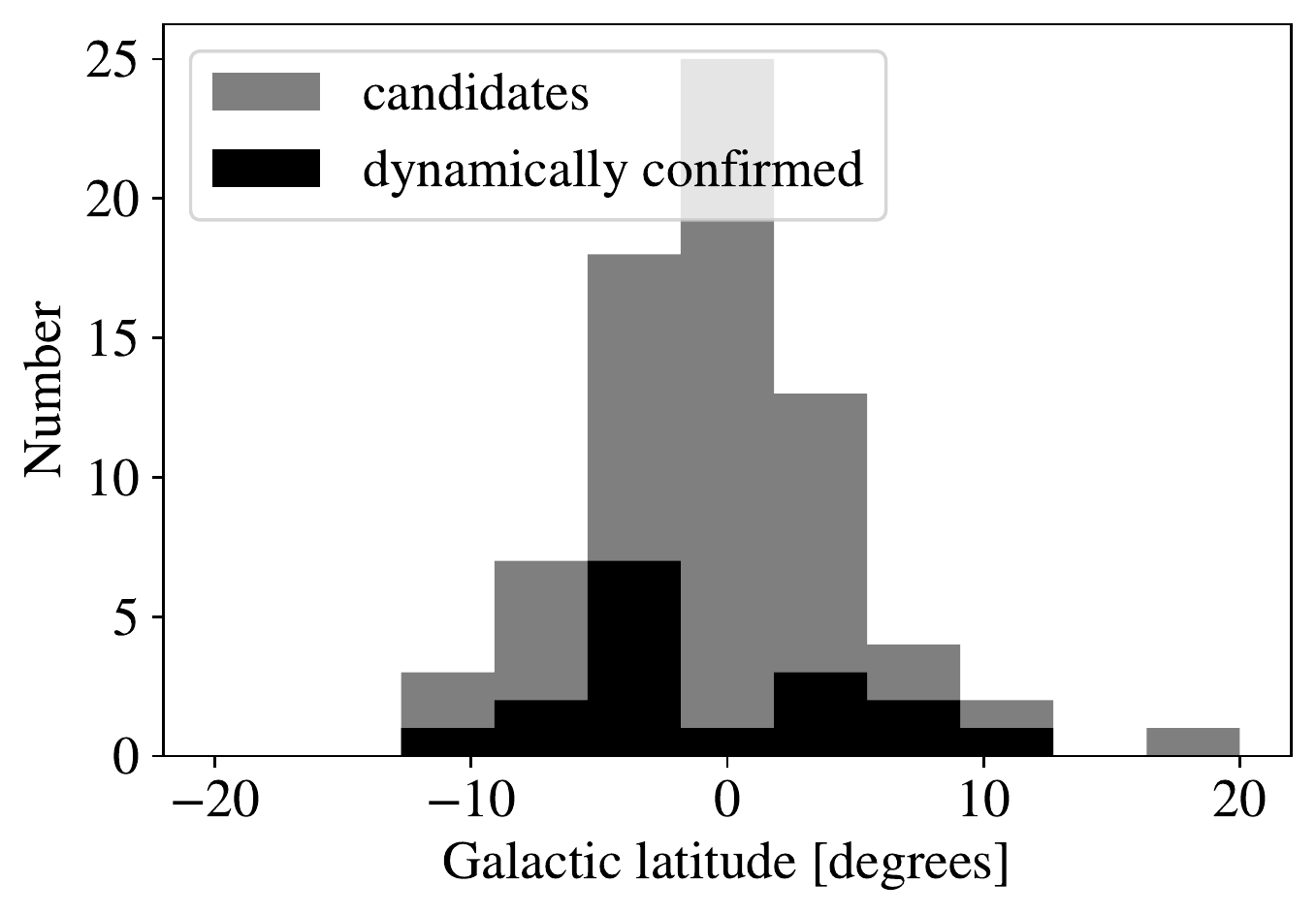}
    \caption{The angular distribution of BHs and candidates in X-ray binaries relative to the Galactic midplane. This figure does not include several sources at high Galactic latitudes. The majority of sources are distributed within $\pm 15$ degrees of the Galactic midplane.  }
    \label{fig:lat}
\end{figure}

\begin{figure}[tbp]
    \centering
    \includegraphics[width=0.5\textwidth]{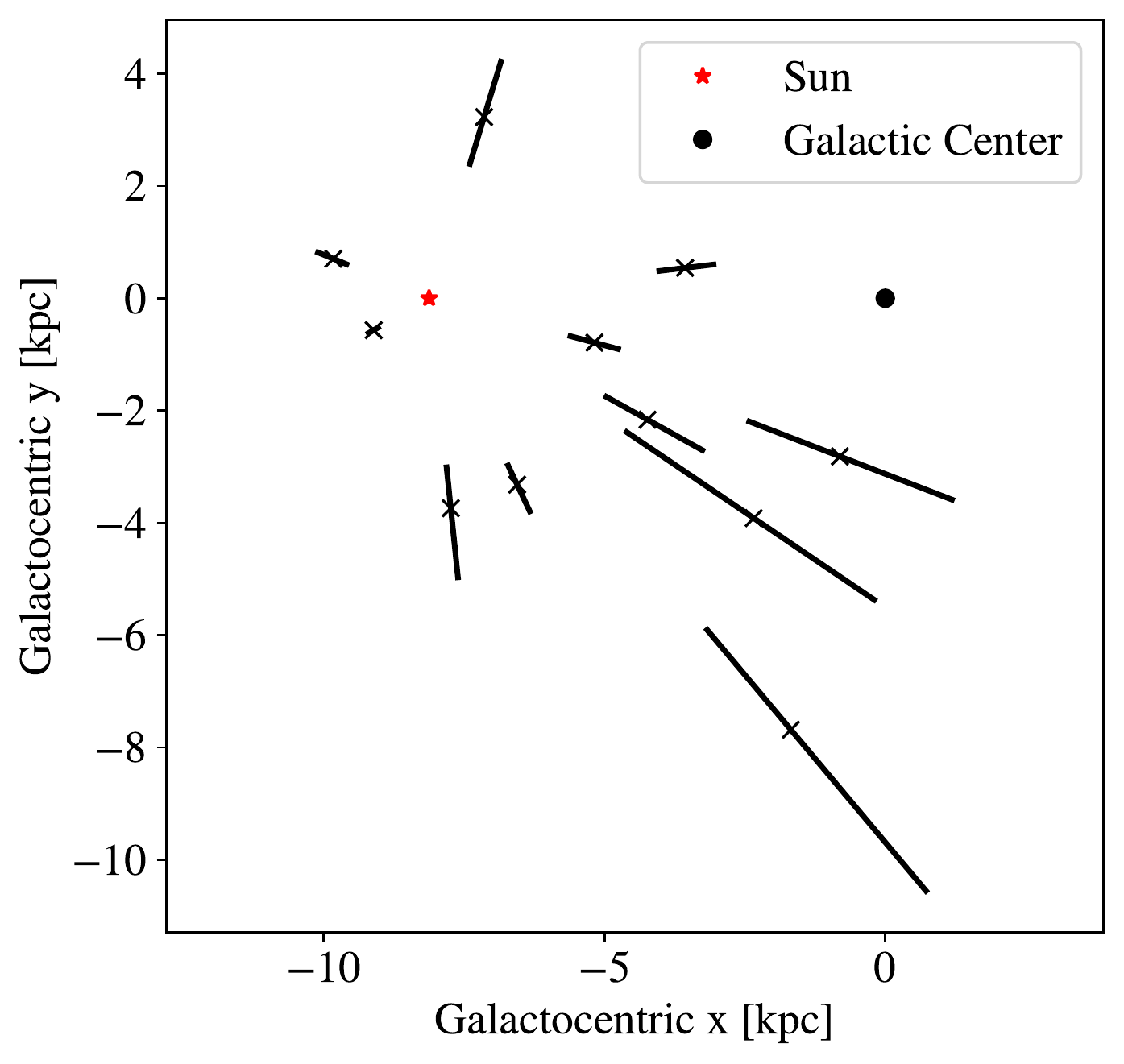}
    \includegraphics[width=0.5\textwidth]{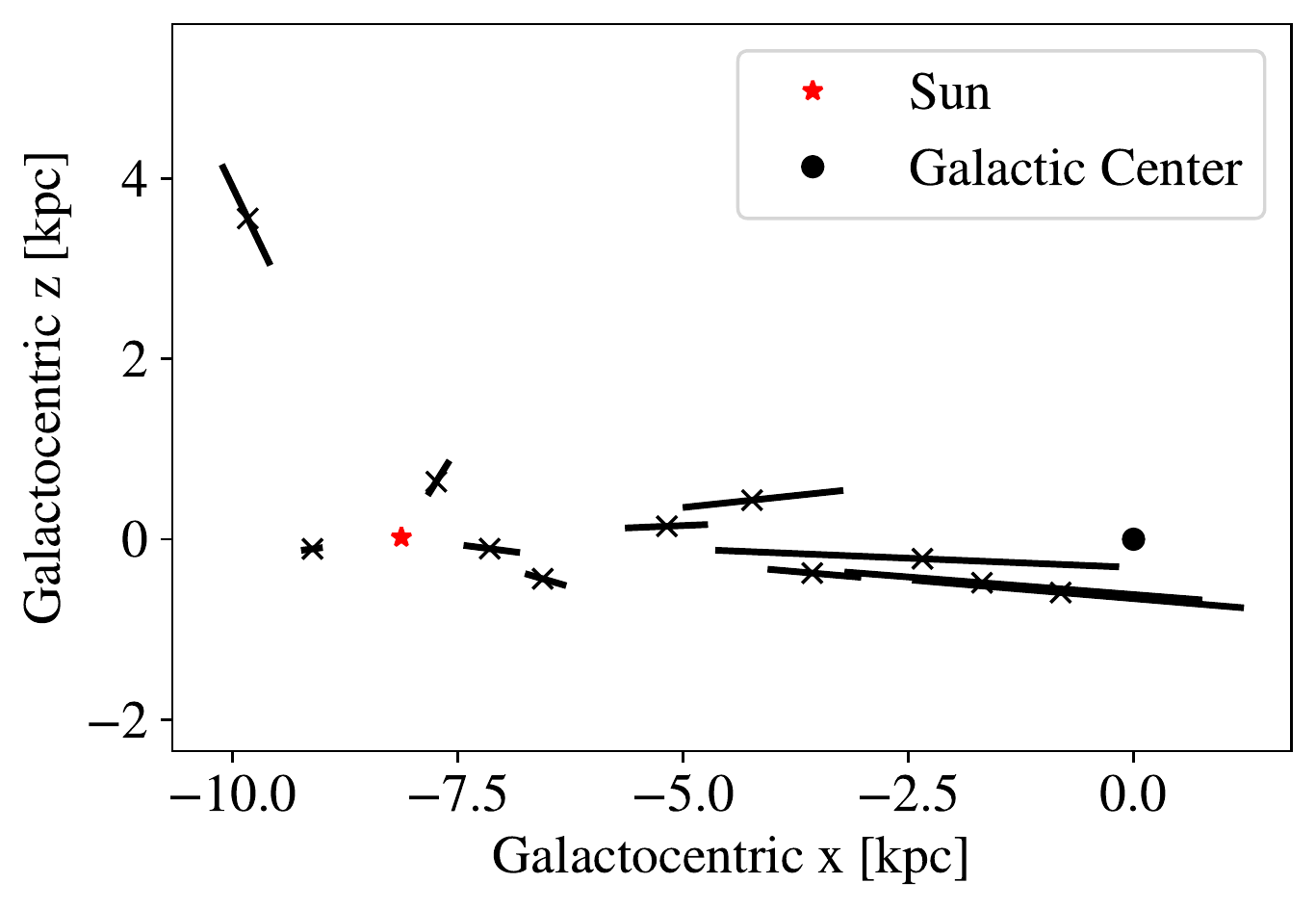}
    \includegraphics[width=0.5\textwidth]{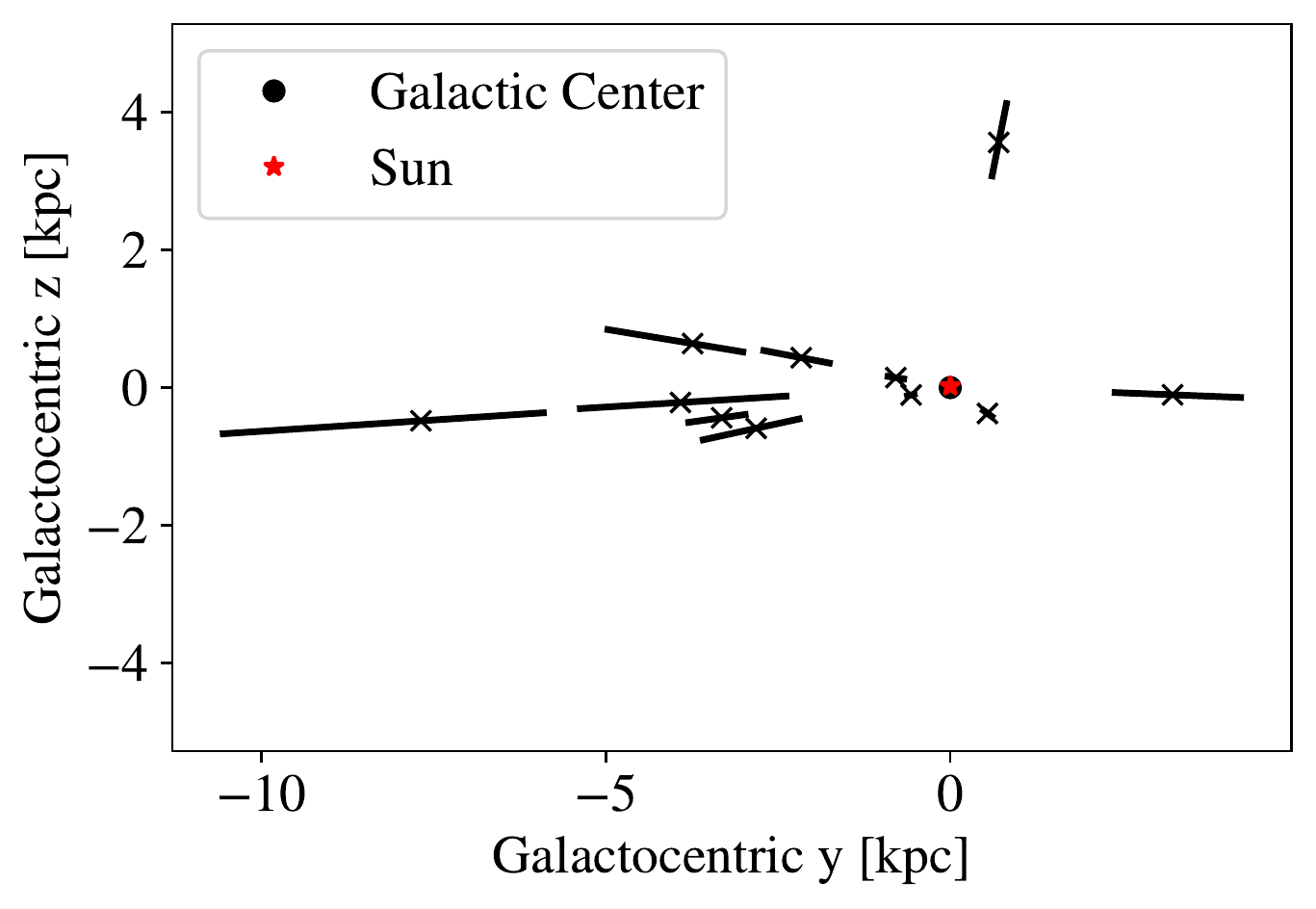}
    \caption{Three-dimensional, galactocentric locations of dynamically confirmed BH-LMXBs with Gaia counterparts. These figures use the photogeometric distance estimate, and lines show the associated distance uncertainty. The Gaia-detected sources cluster in, roughly, the solar quadrant of the Galactic disk. Many lie within $\pm 1$ kpc in $z$, but at least one source is elevated to lie within the Galactic halo.}
    \label{fig:Gaia}
\end{figure}

Figure \ref{fig:skydist} shows the sky distribution of Galactic BH X-ray binaries from the WatchDog catalog \cite{2016ApJS..222...15T} in Galactic coordinates. Dynamically confirmed BHs and their candidates extend along the Galactic plane, with a few notable exceptions at higher latitudes. Some of these objects have counterparts in Gaia DR3, and are highlighted in Figure \ref{fig:skydist}. Examination of this sky distribution immediately reveals that the BH X-ray binaries form a ``thick disk" population within the Galaxy. Their distribution, especially that of the candidate objects, is concentrated toward the bulge but extends across Galactic longitude. The vast majority of the sources are within $\pm$15 degrees of the Galactic midplane (Figures 
\ref{fig:lat} and \ref{fig:Gaia}). It is important to at least consider the possible role of selection effects in shaping this distribution. Jonker et al.\cite{2021ApJ...921..131J} argue that the lower line-of-sight extinction outside of the thin disk biases us to observe systems at large scale height. 

One of the conclusions that has been drawn from this sky distribution is the relative role of kicks to newly-formed BHs. In particular, Nelemans et al.\cite{1999A&A...352L..87N} considered the center-of-mass motion imparted by mass loss in supernovae explosions, and showed that the ejecta mass can be related to the Galactic orbits of the X-ray binaries, under the assumption of a symmetrical explosion. Subsequent work revisited these conclusions with new data \cite{2004MNRAS.354..355J,2012MNRAS.425.2799R,2015MNRAS.453.3341R,2017MNRAS.467..298R}.   Repetto et al.\cite{2017MNRAS.467..298R} came to the conclusion that at least some of the BH binaries -- those furthest from the Galactic midplane -- must have received natal kicks. 

As data from the Gaia mission has supplemented our understanding of the distances to X-ray binaries through detections of their donor-star companions\cite{2019MNRAS.485.2642G}, a picture that necessitates relatively strong BH kicks has emerged. 
Atri et al.\cite{2019MNRAS.489.3116A} used 6-dimensional kinematics to show that for many systems, BH natal kicks of $\gtrsim70$km s$^{-1}$ are required. Gandhi et al.\cite{2020MNRAS.496L..22G} have shown that the systems furthest from the Galactic midplane have the shortest orbital periods, as would be expected since these compact systems would be most likely to retain their BH despite a large kick magnitude.

\subsection{Formation of BH-LMXBs}

The formation history of BH-LMXBs is currently only weakly constrained by their observed properties. Particularly puzzling is that many (14 out of 19) share remarkably similar companion star properties -- of G to M dwarfs. Since these stars are likely main sequence stars of mass less than a solar mass, it is puzzling how they come to be close companions to the BH. For example, the pre-main sequence lifetime of a 1 solar mass star is of similar length or longer than the few million years it takes a massive star to evolve and leave behind a BH remnant.  We mention here some theoretical models describing the possible histories of these objects, with a focus on the testable predictions of these scenarios. 

\subsubsection{Common Envelope Model}
The originally-proposed model for the formation of A0620-00 involved a very asymmetric binary system ($M_1\approx 40M_\odot$, $M_2\approx1M_\odot$) in a wide orbit of hundreds of days\cite{1987A&A...183...47D}. The system was then believed to tighten significantly during a common envelope phase\cite{1976IAUS...73...75P}, which would also strip the hydrogen envelope of the massive star, leaving its helium core. As the helium core undergoes further nuclear evolution, it eventually collapses to a BH, leaving a BH main sequence binary with an orbital period of less than a day. From there, angular momentum loss to a combination of tidal dissipation, stellar evolution, magnetic braking, and gravitational waves might drive the system toward the Roche limit where mass transfer from the main sequence star onto the BH can commence\cite{1987A&A...183...47D,2002ApJ...565.1107P}. 

This model, with variations of the initial parameters, might apply to the entire variety of BH-LMXB systems\cite{2002ApJ...565.1107P}. The basic premise of the common envelope formation scenario is that orbital tightening comes at the expense of the binding energy of the massive star envelope\cite{1993PASP..105.1373I,2013A&ARv..21...59I} $\Delta E_{\rm orb} \sim -\Delta E_{\rm env}$. Given uncertainties in massive star structure, and more pressingly, in the exact core--envelope distinction\cite{1994MNRAS.270..121H,2001A&A...369..170T}, estimating the envelope binding energy, especially in massive stars, remains a very active area of research\cite{2016A&A...596A..58K,2021A&A...645A..54K,2022MNRAS.511.2326V}.

In principle, the orbital separation (and thus orbital energy) can change by orders of magnitude from the outset to the conclusion of a common envelope phase\cite{1999ApJ...521..723K}. There is, however, a minimum final separation based on the requirement that the companion not be disrupted by the helium core, resulting in a complete merger of the pair. If the envelope is to be expelled, enough energy must be input while the binary remains at sufficiently wide separation to avoid merger. Recently, Kruckow et al.\cite{2016A&A...596A..58K} have estimated the minimum companion mass that could eject the envelope of massive star models computed with a one-dimensional stellar evolution code. Their findings suggest that for a $1M_\odot$ star to eject the envelope of a massive star, that star would have to be very near its maximal radius $R_1\gtrsim10^3R_\odot$, such that its envelope is as weakly-bound as possible. In practice, these interactions likely do occur binaries appear to have broad distributions of initial orbital properties. However, whether this is too restrictive a condition on the initial phase space remains to be seen\cite{2021A&A...645A..54K}. 

A second puzzle of the common envelope formation scenario is the apparent similarity of the BH-LMXB secondary objects. Their relatively uniform low masses (of similar to a solar mass or less) is not in clear agreement with the common envelope formation model. Indeed, since binaries start with a range of properties of mass ratios and orbits, we might expect the mass transferring population would share some broad distribution -- perhaps even weighting more strongly toward the more-massive secondaries that might more easily eject the primary-star's envelope\cite{2013A&ARv..21...59I}. A possible resolution to this particular concern comes in theoretical models of the mass-transferring system's lifetimes, which are expected to steeply decline as the secondary becomes more massive, as described in what follows and by Podsiadlowski et al.\cite{2002ApJ...565.1107P}. This might imply that BH-LMXBs could have assembled with more massive secondaries that have been reduced over time by mass transfer. 

In summary, there remain broad uncertainties and some possible tensions between current theories of common envelope ejection and the existence of the BH-LMXB systems\cite{1999ApJ...521..723K}. These questions have prompted research into alternative formation mechanisms, which we discuss in what follows. 

\subsubsection{Tidal Capture Formation of BH-LMXBs in Wide Binaries}

Given the tensions associated with the application of the common envelope model to the formation of the BH-LMXBs, alternative options have been explored in the literature. These generally combine a form of dynamical scatterings combined with the tidal capture of very eccentric BH--star orbits into circular, mass-transferring binaries\cite{2016MNRAS.458.4188M,2017MNRAS.469.3088K}. 

One such theory involves scatterings occurring naturally in the wide binaries distributed  throughout the milky way disk. In this case, a BH, either single or in a binary system, interacts with the surrounding stars in its environment. The most prevalent objects are low-mass main sequence stars and their binaries (approximately 40\% of low-mass main sequence stars are in binaries with an approximately flat distribution of log period from $10^2$ to $10^8$~d, see figures 37 and 38 of Moe et al.\cite{2017ApJS..230...15M}. 

The interaction cross section between a binary and another object (with impact parameter $<b$) is
\begin{equation}
\sigma \approx \pi b^2 \left( 1+ \frac{2 G M_{\rm tot}}{b v_\infty^2} \right), 
\end{equation}
where $b$ is the impact parameter, $M_{\rm tot}$ is the total mass of the objects, and $v_\infty$ is the relative encounter velocity, which, in an isotropic stellar distribution is similar to the three-dimensional velocity dispersion.  Therefore, encounters in which $b< 2G M_{\rm tot} / v_\infty^2$ are in the gravitational-focus dominated regime, in which $\sigma \propto b$. Otherwise, when $b>2G M_{\rm tot} / v_\infty^2$, the geometric cross section dominates and $\sigma \propto b^2$.  The corresponding interaction rate per object is given by 
\begin{equation}
\Gamma \approx n_\ast \sigma v_\infty
\end{equation}
where $n_\ast$ is the number density of surrounding stars. 

If we imagine the population of very wide binaries (those with semi-major axes $a\sim 10^4$~AU) as the typical target, encounters with similar or smaller impact parameter occur at a rate, 
\begin{equation}\label{rate}
\Gamma \approx 3\times10^{-8}~{\rm yr}^{-1} \left( \frac{n_\ast}{0.1 {\rm pc}^{-3}} \right)  \left( \frac{b}{10^4{\rm AU}} \right)^2 \left( \frac{v_\infty}{40{\rm km~s}^{-1}} \right)  ,
\end{equation} 
where $n_\ast$ is the local stellar number density. The expression above implies that over a Hubble time, a BH  in the Galactic field will experience several hundred encounters with $b\lesssim 10^4$~AU.

These binary--single or binary--binary encounters can cause the BH to gain or exchange a wide binary partner. They also randomize the angular momentum of the BH -- star system. We can estimate the fraction of the randomized orbital phase space that leads to tidal capture outcomes as follows. The requirement that the orbit circularize within a few stellar radii suggests that the original periapse distance was $r_{\rm p,cap} \sim a_{\rm circ}/2 \sim f_p R_\ast$, where $f_p$ is a factor on the order of a few and given that tides have minimal effect on the orbital angular momentum \cite{1975MNRAS.172P..15F,1977ApJ...213..183P}. If systems have randomized -- or thermal -- distributions of angular momentum the fraction of orbits with periapse distance less than $r_{\rm p,cap}$ is approximately $r_{\rm p,cap}/a$. For $a\sim 10^4$~AU and $r_{\rm p,cap}\sim10R_\odot$, we find $r_{\rm p,cap}/a \sim 5\times 10^{-6}$. If encounters with small enough $b$ efficiently refill this phase space, then on the order of that fraction of systems will experience a tidal capture every orbital period scattering. Associating $b\sim a$, the tidal capture rate becomes,
\begin{equation}
\Gamma_{\rm cap} \approx 10^{-13}~{\rm yr}^{-1} \left( \frac{n_\ast}{0.1 {\rm pc}^{-3}} \right)  \left(\frac{r_{\rm p,cap}}{10R_\odot} \right) \left( \frac{a}{10^4{\rm AU}} \right) \left( \frac{v_\infty}{40{\rm km~s}^{-1}} \right).
\end{equation} 
This implies that on the order of one in a thousand wide binaries with a BH will experience a tidal-capture forming encounter in a Hubble time. Thus a base population as small as $10^5$ wide BH binaries could produce a population of $10^2$ BH-LMXBs over a Hubble time. The estimate here ignores some subtleties of the scattering and capture process in favor of an order of magnitude estimate, several recent papers\cite{2014ApJ...782...60K,2016MNRAS.458.4188M,2017MNRAS.469.3088K} provide a more detailed discussion of the dynamics. 

We emphasize two considerations that can modify the conclusions of the dynamical models\cite{2016MNRAS.458.4188M,2017MNRAS.469.3088K}. If BHs receive natal kicks, their ability to retain very wide binary companions or to capture these very wide companions in binary--single interactions is significantly restricted. Because the velocity at infinity is much larger than the orbital velocity, the binary is essentially at rest as the BH passes, and the encounter becomes impulsive, rather than chaotic. Then the probability of the BH capturing a companion is limited by its gravitational capture cross section. For a single, $10 M_\odot$ BH moving relative to the background field stars (including wide binaries) at $v_\infty = 40$~km~s$^{-1}$, the gravitational capture cross section is limited to $b\lesssim 2 G M_{\rm bh}/v_\infty^2 \approx 0.1$~AU. Examining  equation \eqref{rate} shows that these encounters are rare and thus highly unlikely. Secondly, the discovery by Gandhi et al.\cite{2020MNRAS.496L..22G} that the BHs furthest from the orbital midplane have the tightest companions. A priori this appears consistent only with a picture in which the BHs were already bound to their companions at the time that the kick was received -- and not that they were captured later from the field.  

\subsubsection{Tidal Capture formation of BH-LMXBs in Hierarchical Multiples}

A related avenue for BH-LMXB formation that has been suggested similarly combines dynamical rearrangement of orbits with tidal capture of a star around a BH. In this case, we instead focus on hierarchical systems that contain a BH and a lower-mass main sequence star in an inner binary, with a tertiary in an inclined outer orbit. Under certain conditions these systems undergo secular oscillations between the inclination and angular momentum of the inner and outer orbits (known as Kozai-Lidov oscillations). These oscillations can be terminated if the inner orbit becomes so highly eccentric that tidal dissipation causes the inner orbit to capture into a much more compact, circular configuration, perhaps leading to the formation of a BH-LMXB\cite{2016ApJ...822L..24N}. A small fraction of systems enter into chaotic phases, possibly with a similar, highly eccentric fate \cite{2022A&A...661A..61T}. Many of the Naoz et al.\cite{2016ApJ...822L..24N} numerically-simulated systems reach very eccentric configurations or cross the Roche limit while still eccentric. Thus, to a crucial extent, the fate of these systems (and whether or not they eventually form stable, circular BH-LMXBs) depends on the result of these very strong tidal interactions and the nonlinear tidal dissipation that likely results \cite{2004ApJ...601L.171K,1987A&A...184..164R,2022ApJ...937...37M}. 

While the Naoz et al.\cite{2016ApJ...822L..24N} models neglected the possible role of BH natal kicks, the update of their results in the systematic study of Lu et al.\cite{2019MNRAS.484.1506L}, is strongly suggestive that the reconfiguration of systems due to natal kicks still allows for the formation of many BH-LMXBs through this channel. Much like the field dynamical channel, it is unclear whether this channel can produce the observed period versus Galactic scale height anticorrelation of Gandhi et al.\cite{2020MNRAS.496L..22G}.

\subsection{Lifetime of BH-LMXBs}
Tetarenko et al.\cite{2016ApJS..222...15T} have collated estimates of the mass exchange rate in BH binaries given their X-ray luminosities (see Figure \ref{fig:mdot}). These estimates vary widely between systems, in the range of $\dot M\sim 10^{-12}$--$10^{-8} M_\odot$~yr$^{-1}$. For donor stars on the order of a solar mass, the estimated lifetime $M/\dot M$, can easily be a Gyr or more. Evidently the BH-LMXBs can be a very long-lived population once assembled. 

We further clarify these estimates by comparison of the equilibrium mass transfer rates driven by gravitational wave losses\cite{2002ApJ...565.1107P,2015ApJ...800...17F} for fully convective donor stars, with $R_{\rm donor} \propto M_{\rm donor}^{-1/3}$ response to mass loss. We illustrate the behavior of three example mass-transferring systems with differing initial donor mass of 0.5, 1.0, and 1.5$M_\odot$, each paired with a $7M_\odot$ BH accretor. In these systems, the equilibrium mass transfer rate is a strong function of the instantaneous mass of the donor. The more massive donors lose their mass rapidly, and the systems converge, joining similar evolutionary tracks regardless of the original mass. For example, after 0.1~Gyr, the 1.5 and 1$M_\odot$ models each have reduced in donor mass to approximately $0.75M_\odot$. These models show that the donor stars tend to converge on solar or subsolar masses, regardless of initial mass, perhaps going some way toward explaining the broadly uniform low masses of BH-LMXB donor stars. Secondly, in accreting from the donors in these systems, the BH mass increases by on the order of a solar mass, sufficient to generate high dimensionless spins\cite{2015ApJ...800...17F}. 

\begin{figure}
    \centering
    \includegraphics[width=0.5\textwidth]{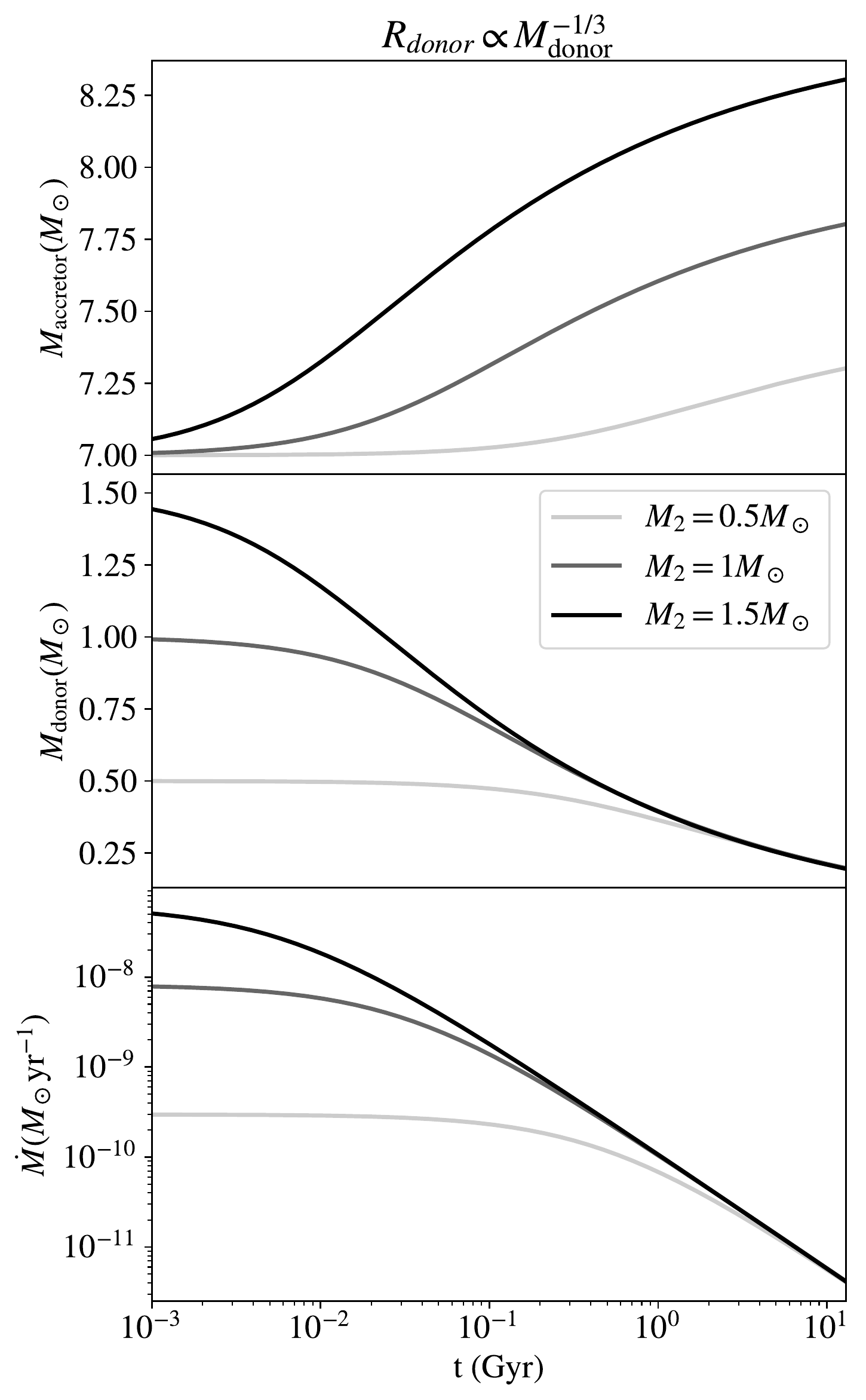}
    \caption{An example mass transfer history of an idealized BH-LMXB system. We imagine a $7M_\odot$ BH accreting from donors of 0.5, 1.0, and 1.5$M_\odot$ initial mass. The more rapid the donor, the more rapid the mass loss, such that the systems converge in donor masses at ages greater than approximately 100 Myr, perhaps helping to explain the similarities of BH-LMXB donors in mass and spectral type.  }
    \label{fig:mt}
\end{figure}

\section{The Spins of Black Holes in X-ray Binaries}\label{sec:spin}

BHs are characterized by two numbers: their mass and spin. So far, we have focused on the easier of these two quantities to ascertain, the BH mass, but progress has also been made toward measuring their spins. Measurements of BH spin depend on the existence of an innermost stable circular orbit (ISCO) at radius $r_{\rm isco}$. For a non-spinning BH, $r_{\rm isco}=6 GM/c^2$, while for a maximally spinning BH orbits down to $GM/c^2$ can exist aligned with the spin vector. 

Thus, if X-ray data can measure the innermost radius of an accretion disk and separately constrain the BH mass, we are in a position to measure $r_{\rm isco}$ in units of $GM/c^2$. Two methods have been explored to date, the first, continuum fitting, is based on fitting the spectral energy distribution of thermal emission from an accretion disk surrounding the BH, with the key concept that a high-energy cutoff occurs at the ISCO, where the gravitational potential energy is greatest, and that no emission comes from $r<r_{\rm isco}$.\cite{2011CQGra..28k4009M,2014SSRv..183..295M}  A second method is based on resolving the Fe K$\alpha$ line in accretion spectra. Its width will be Doppler broadened by the fastest motions in the accretion disk, which occur at the smallest radii.\cite{1989MNRAS.238..729F,2011CQGra..28k4009M,2014SSRv..183..277R}  

\begin{figure}
    \centering
    \includegraphics[width=0.49\textwidth]{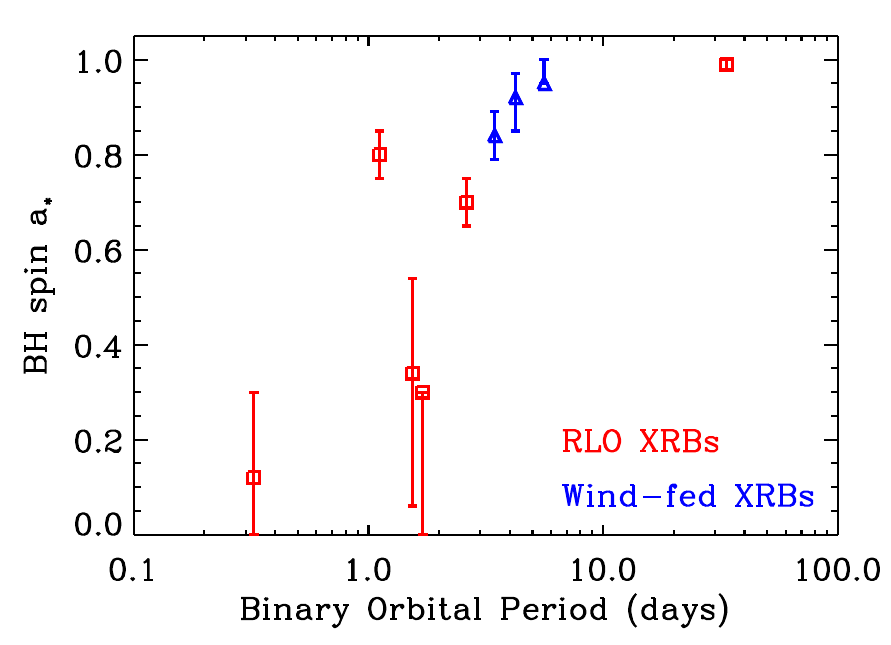}
    \includegraphics[width=0.49\textwidth]{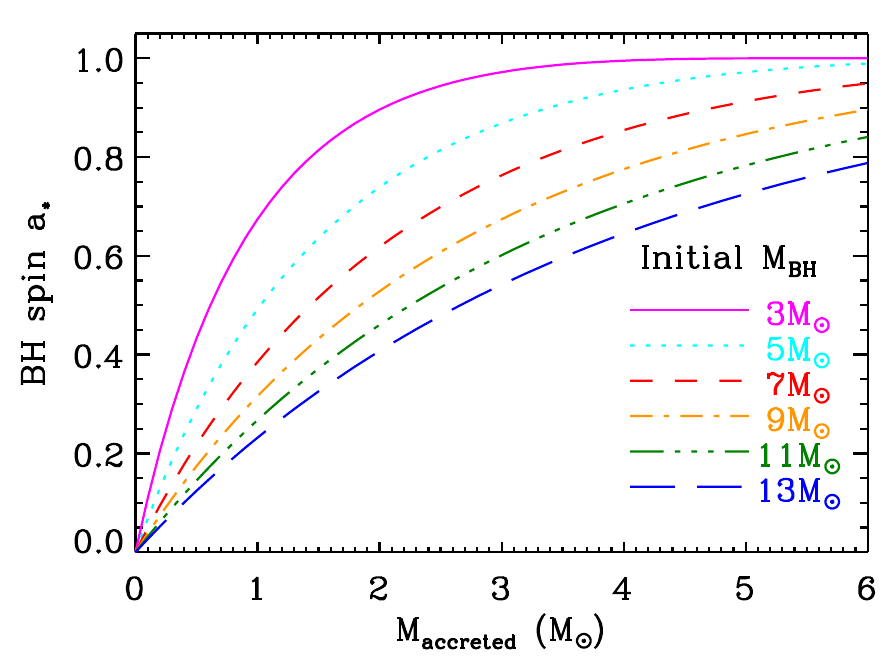}
    \caption{Measured spins in BH X-ray binaries (left) and the acquired spin due to accretion from an ISCO (right). BHs exhibit a correlation between spin and orbital period. They may gain some of this spin through their accretion. Figure reproduced with permission from Fragos and McClintock in 2015\cite{2015ApJ...800...17F}. }
    \label{fig:spin}
\end{figure}

One recent summary of measured BH spins is presented by Reynolds\cite{2021ARA&A..59..117R}. Measured BH spins range from low to extremely high values, and appear to be correlated with system orbital period. Figure \ref{fig:spin} shows spins of BHs tabulated by Fragos and McClintock in 2015\cite{2015ApJ...800...17F}. Shorter orbital period systems tend to correspond to lower BH spins. By contrast some of the longer orbital period systems have dimensionless spins $a$ near unity.  The right panel of Figure \ref{fig:spin} relates these spins to the ongoing mass transfer and accretion in BH X-ray binaries. When BHs accrete material with angular momentum similar to that of the ISCO, they gain spin along with mass. Trajectories are shown given different initial BH masses at $a=0$. A qualitative conclusion can be drawn from this diagram. If BHs in X ray binaries gain on the order of a solar mass or more due to accretion, they may very likely acquire significant spin as well through this process. 

\section{Non X-ray Emitting Galactic Black Holes}\label{sec:nonxray}

In addition to the X-ray emitting systems discussed so far, several BHs or BH candidates have been detected through their gravitational effect on their surroundings. 

\subsection{Gravitational Microlensing}
The gravitational microlensing technique was proposed as a means of detecting dark mass along lines of sight to dense stellar fields.\cite{1986ApJ...304....1P,1991ApJ...371L..63P} The method is based on time domain monitoring in the optical. Here surveys attempt to detect the signatures of magnification of stellar light when an intervening object passes between the observer and the source star due to the proper motions and parallaxes of source and lens objects on the sky. Because the distribution of stars in the Galaxy is generally quite sparse, these searches have targeted areas where the source stars are as dense as possible.\cite{1986ApJ...304....1P,1991ApJ...371L..63P} For example, the MACHO survey\cite{1993ASPC...43..291A} targeted the site lines through the Galactic halo to the Large Magellanic cloud and the Galactic bulge. The current generation of the OGLE microlensing survey covers the Galactic disk, bulge, and fields surrounding the Magellanic  clouds\cite{2019ApJS..244...29M}. 

The apparent brightening that a source experiences due to an intervening lens is related to the angular distance between the source and the lens. The total magnification factor can be expressed
\begin{equation}
    A = \frac{u^2 + 2 }{u(u^2+4)^{1/2}},
\end{equation}
where $u=r/R_0$, and $r$ is the projected distance between the source and the lensing mass. The normalizing distance $R_0$ is the Einstein radius,
\begin{equation}
    R_0^2 = \frac{4 GMD}{c^2},
\end{equation}
where 
\begin{equation}
    D = \frac{D_l (D_s - D_l)} {D_s},
\end{equation}
where $D_l$ is the distance to the lens and $D_s$ is the distance to the source\cite{1986ApJ...304....1P,1991ApJ...371L..63P}. If we take the example case of $D_l \approx D_s/2$, then we find $D\approx D_s/4$ and $R_0^2\approx GMD_s/c^2$. Solar mass objects in the Galactic bulge thus have Einstein radii on the order of 1 mas.\cite{1991ApJ...371L..63P} Finally, the magnifcation factor of the source relates to the magnitude increase by $\Delta$mag$=2.5 \log_{10}\left( A \right)$. 

Given an angular size of the Einstein radius for a given source-lens pair, the probability that a lens crosses a source may be computed. We call this quantity the optical depth to lensing. This quantity is directly related to the density of lens objects on the sky. Taking the density of lenses to be the stellar density in the disk along the line of sight to the Galactic bulge, the optical depth is on the order of\cite{1991ApJ...371L..63P} $\tau_{\rm lens} \sim 10^{-6}$. The optical depth toward the Magellanic clouds was predicted to be similar if the dark matter halo of the Galaxy were composed of stellar mass dark objects like BHs\cite{1986ApJ...304....1P}. These low optical depths imply that, to discovery microlensing events efficiently, surveys need to monitor millions of potential source stars. 

\begin{figure}
    \centering
\includegraphics[width=0.9\textwidth]{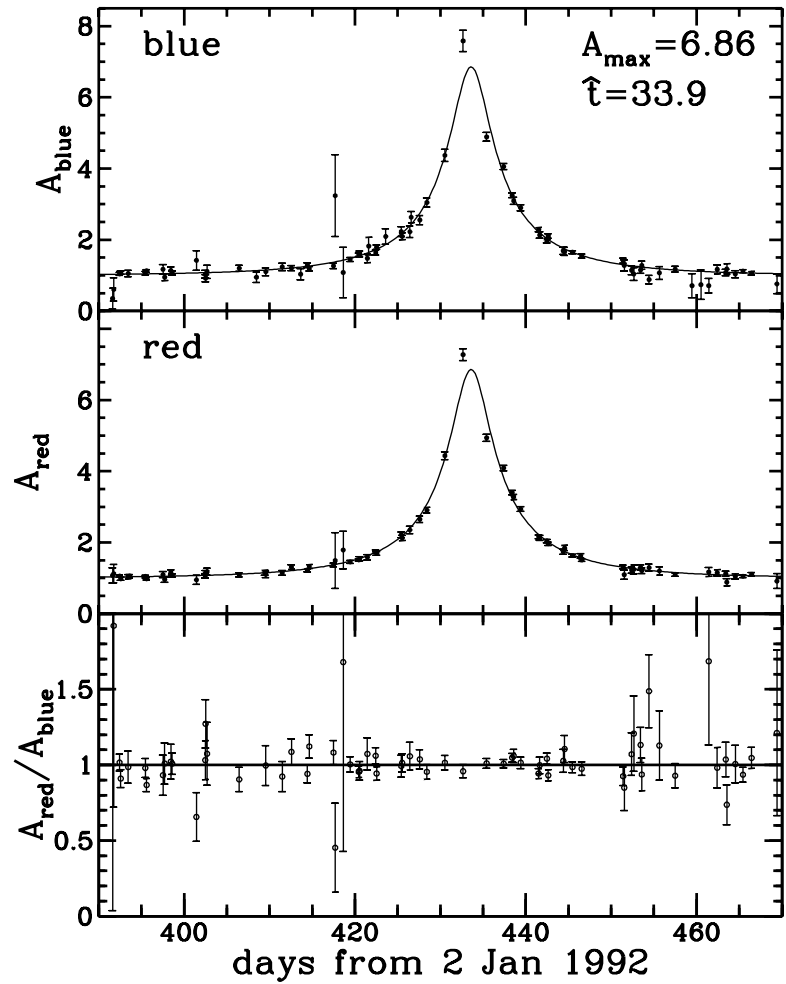}
    \caption{The first known gravitational microlensing event in the Galactic halo discovered by the MACHO survey. A key feature of microlensing is the achromatic photometric amplification. Figure reproduced with permission from Alcock et al.\cite{1993Natur.365..621A} }
    \label{fig:microlens}
\end{figure}

Figure \ref{fig:microlens} shows the discovery by reported in 1993 by Alcock et al. and the MACHO survey project of a microlensing event in the direction of the Large Magellenic Cloud\cite{1993Natur.365..621A}. A key characteristic is that the light curve does not exhibit chromatic variation, even as the magnification factor changes greatly. This is because (in the limit that that angular size of the source is smaller than the Einstein radius) $A$ depends solely on the geometry of the source and lens, and not on the wavelength.

\begin{figure}
    \centering
\includegraphics[width=0.9\textwidth]{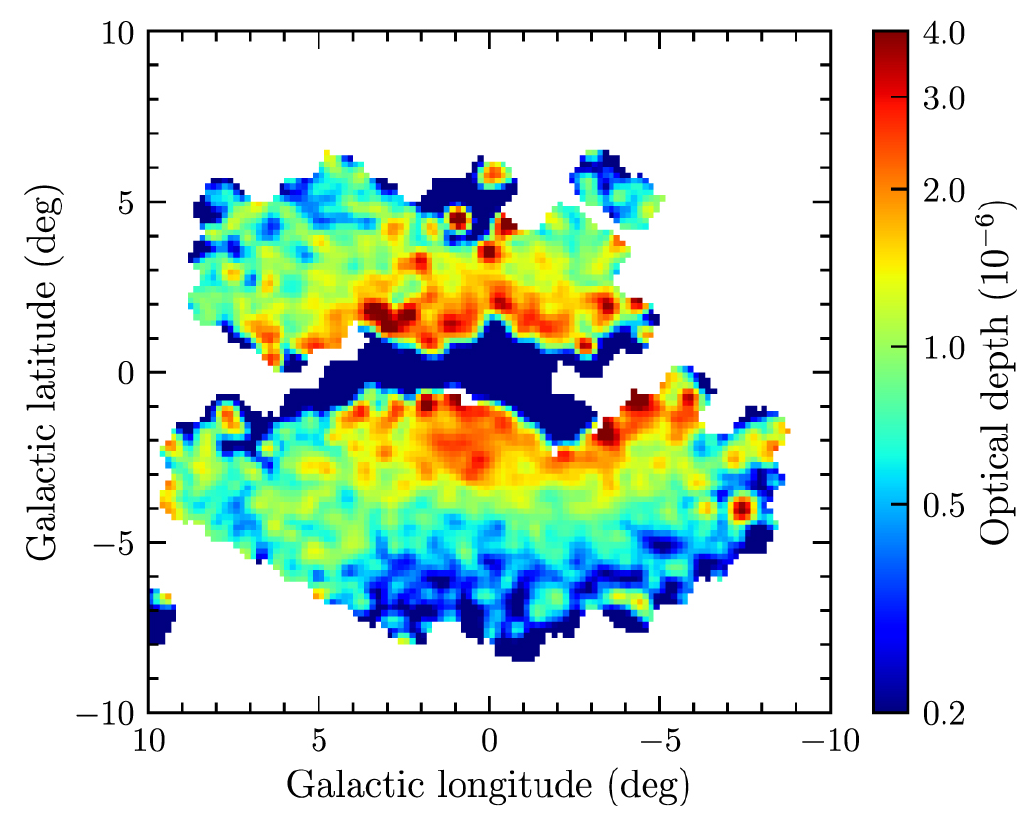}
    \caption{Optical depth to microlensing, as derived from the OGLE-IV survey. Figure reproduced  with permission from Mr\'oz et al.\cite{2019ApJS..244...29M}. }
    \label{fig:taulens}
\end{figure}

In spite of the low lensing optical depths, the discovery of similar microlensing effects has become commonplace. Figure \ref{fig:taulens} shows the cumulative optical depth to  microlensing events derived from the OGLE-IV survey by Mr\'oz et al.\cite{2019ApJS..244...29M}. As predicted\cite{1991ApJ...371L..63P}, the lensing optical depth is $\sim 10^{-6}$, and represents the lensing by stars, compact objects, brown dwarfs, and planets. 

Through these detections, microlensing surveys have placed important constraints on the portion of dark matter that can be made up of stellar-mass compact objects. For example, the MACHO survey ruled out a 100\% contribution of stellar mass objects to the Milky Way's dark matter halo at the 95\% confidence level following the first 6 years of observations\cite{2000ApJ...542..281A}. More recently, Mr\'oz et al.\cite{2019ApJS..244...29M} discuss the consistency between the OGLE-IV lensing optical depth and the distribution of luminous sources like stars in the Galactic disk. Thus, one conclusion of these extensive surveys is that stellar-mass BHs cannot fully explain the Milky Way's dark matter budget.

\subsection{Astrometric Microlensing and Black Holes}

\begin{figure}
    \centering
\includegraphics[width=0.9\textwidth]{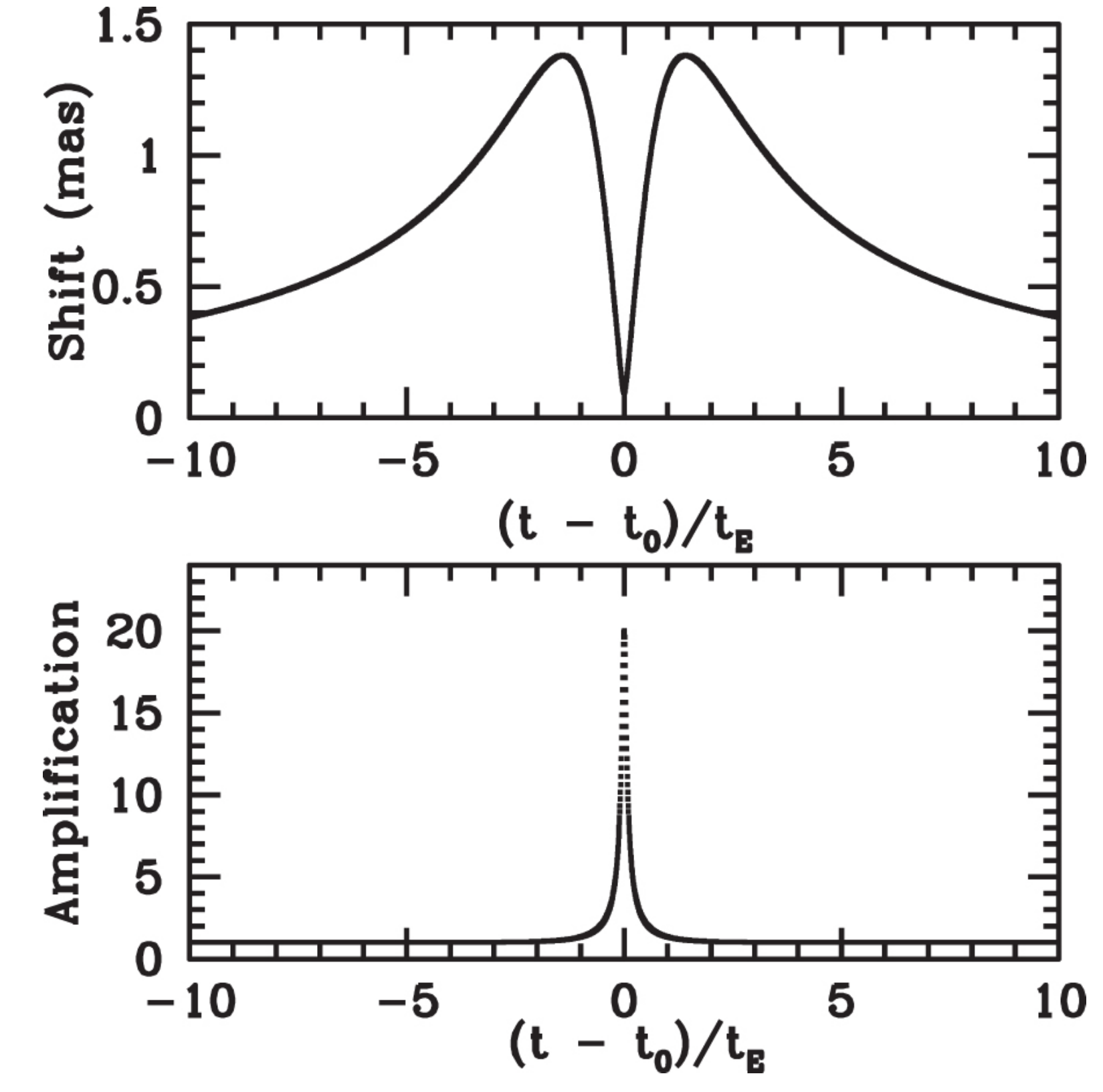}
    \caption{Astrometric shift (top panel) and photometric amplification (bottom panel) as a function of time in terms of Einstein radius crossing times, $t_E$. Astrometric shifts are on the order of the Einstein radius of the lens, and have long duration relative to the photometric signature. This example assumes a $5M_\odot$ BH at 2~kpc passing in front of a source star at 8~kpc. Figure reproduced with permission from Sahu et al.\cite{2022ApJ...933...83S}.}
    \label{fig:astroBH}
\end{figure}

Detections of photometric microlensing can constrain the total number of stellar-mass lens objects, but they do not, in general identify the nature of the lens of object. There is, for example, always degeneracy between the lens mass and the impact parameter between the source and the lens.\cite{2002ApJ...576L.131A,2002ApJ...579..639B} Given typical assumptions about the Milky Way's initial mass function and the compact objects that form as a result, BHs may account for on the order of 1\% of the microlensing events. Thus, many BH microlenses are likely to have been detected.\cite{2002ApJ...576L.131A,2016MNRAS.458.3012W,2022ApJ...933L..23L} However, if both photometric magnification and astrometric distortion associated with a lens are observed, we are able to disentangle the source geometry and mass\cite{2000ApJ...535..928G,2022ApJ...930..159A}. For a detailed review of these effects see Sahu et al.\cite{2022ApJ...933...83S}. Figure \ref{fig:astroBH} shows a hypothetical combined photometric and astrometric signature from an isolated BH lensing a Galactic bulge source star.

Beginning in 2022, Hubble Space Telescope observations of several microlensing BH candidates have been used to place joint astrometric and photometric constraints on their properties.\cite{2022ApJ...933L..23L,2022ApJS..260...55L,2022ApJ...937L..24M,2022ApJ...933...83S} These BH candidates have masses that are estimated through the combination of photometry and astrometry and masses that place them above the mass of other potential compact objects like white dwarfs and neutron stars. 

Sahu et al. report the first microlensing discovery of an isolated BH candidate, MOA-2011-BLG-191/OGLE-2011-BLG-0462, with mass constrained through joint photometric and astrometric data\cite{2022ApJ...933...83S}. The BH in this case is the lens, and has a mass of $7.1\pm 1.3M_\odot$ and is located at a distance of approximately 1.6~kpc. The light curve duration constrains the space velocity of the lens, which is offset from the local population by $\sim 45$~km~s$^{-1}$, which Sahu et al. attribute to a potential natal kick of the BH\cite{2022ApJ...933...83S}.  

In a parallel paper, Lam et al.\cite{2022ApJ...933L..23L} analyze the MOA-2011-BLG-191/OGLE-2011-BLG-0462 data and report that the object has a mass of $1.6-4.4M_\odot$, clearly in tension with the measurements of Sahu et al.\cite{2022ApJ...933...83S}. Though both measurements point to a compact object as the lens, the Lam et al. measurement could be consistent with either a NS or low-mass BH. Mr\'oz et al.\cite{2022ApJ...937L..24M} reanalyze the data presented by Sahu et al.\cite{2022ApJ...933...83S} and Lam et al.\cite{2022ApJ...933L..23L} and suggest that background sources are a possible contaminant, and favor a mass of $7.88 \pm 0.82M_\odot$. Thus the final interpertation of this particular object most likely awaits further confirmation. However, these studies clearly demonstrate the value of combined photometric and astrometric microlensing in the search for Galactic BHs.

\subsection{Non X-ray Binaries Containing Black Holes or Candidates}

Another quite recent development has been the discovery of several BHs in binaries not through their X-ray activity, but through their orbital motion. This discovery has been long-sought, and thus there are a number of systems initially identified as candidates and later reinterpreted. Here we focus on reports of several confirmed binaries that likely host BHs. 

Mahy et al.\cite{2022A&A...664A.159M} report on the single-lined spectroscopic binary HD 130298. The system contains an O-star primary, and an unseen companion. The system is observed as a single-line spectroscopic binary and the mass function is constrained. However, the BH mass itself is not directly measured due to an uncertain inclination. Nonetheless, the minimum companion mass for the presumed BH places it squarely in the mass range of a BH, with a lower mass limit of approximately $7M_\odot$. HD 130298's orbital period is approximately 14.6~d. Similarly,  Shenar et al.\cite{2022NatAs...6.1085S} report on VFTS 243, a 25$M_\odot$ O-star in an orbit with a presumed BH of $>9M_\odot$. In this case, the authors argue that a non-degenerate companion is strongly ruled out. The orbital period of the system is 10.4~d.  Each of these systems is a promising equivalent to the known BH-HMXBs, but with a wider orbit and lacking in significant accretion-powered  X-ray emission.

A binary in the globular cluster NGC 3201 is thought to contain a companion of at least $4M_\odot$, as reported by Giesers et al.\cite{2018MNRAS.475L..15G}. Again an unknown inclination makes this a lower-limit rather than a detection. However, the low-mass companion star makes this system a wider analog of known BH-LMXBs. An additional candidate with a minimum mass of 7.7$M_\odot$ was reported in subsequent analysis\cite{2020A&A...635A..65K}. Theoretical modeling of this cluster indicates that it might retain a large number of BHs.\cite{2020A&A...635A..65K}

\begin{figure}[tbp]
    \centering
    \includegraphics[width=\textwidth]{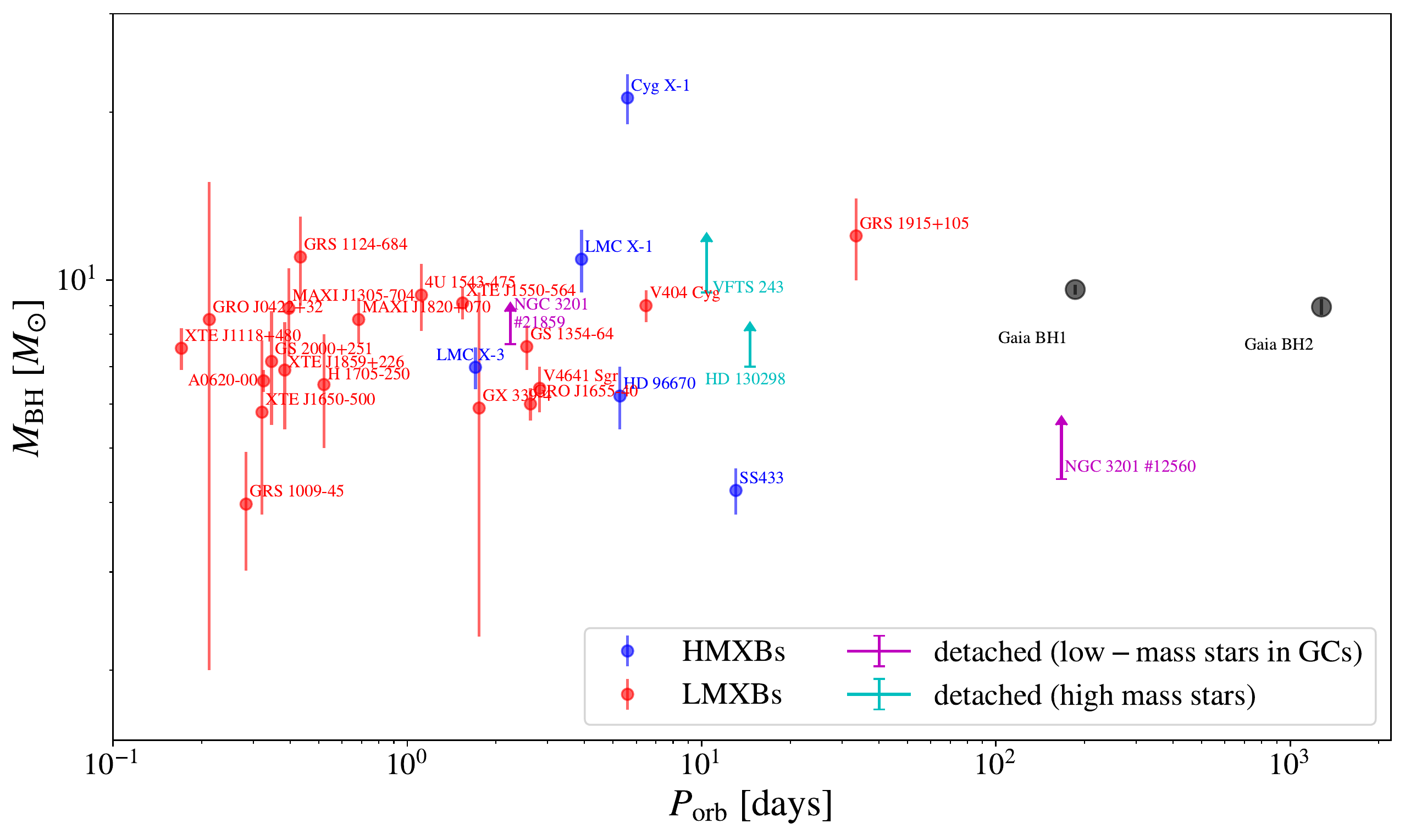}
    \caption{The BH mass versus orbital period space of known BH systems, adapted with permission from El Badry et al.\cite{2023arXiv230207880E} BH masses and periods are shown for the following systems: LMC X-3\cite{2014ApJ...794..154O}, LMC X-1\cite{2009ApJ...697..573O}, Cyg X-1\cite{2021Sci...371.1046M}, SS433\cite{2020A&A...640A..96P}, HD 96670\cite{2021ApJ...913...48G},
    GRO J0422+32\cite{2000MNRAS.317..528W},
    A0620-00\cite{2010ApJ...710.1127C},
    GRS 1009-45\cite{2003ApJ...599.1254G},
    XTE J1118+480\cite{2013AJ....145...21K},
    GRS 1124-684\cite{2016ApJ...825...46W},
    MAXI J1305-704\cite{2021MNRAS.506..581M},
    GS 1354-64\cite{2021MNRAS.506..581M},
    XTE J1550-564\cite{2011ApJ...730...75O},
    4U 1543-475\cite{2002ApJ...568..845O},
    XTE J1650-500\cite{2004ApJ...616..376O},
    GRO J1655-40\cite{2003MNRAS.339.1031S},
    H 1705-250\cite{1997AJ....114.1170H},
    GX 339-4\cite{2017ApJ...846..132H},
    MAXI J1820+070\cite{2020ApJ...893L..37T},
    XTE J1859+226\cite{2022MNRAS.517.1476Y},
    V4641 Sgr\cite{2014ApJ...784....2M},
    GRS 1915+105\cite{2014ApJ...796....2R},
    V404 Cyg\cite{2010ApJ...716.1105K},
    GS 2000+251\cite{2004AJ....127..481I},
    NGC 3201 \#12560 and \#21859\cite{2020A&A...635A..65K},
    VFTS 243\cite{2022NatAs...6.1085S},
    HD 130298\cite{2022A&A...664A.159M},
    Gaia BH1\cite{2023MNRAS.518.1057E}, and
    Gaia BH2\cite{2023arXiv230207880E}. }
    \label{fig:MP}
\end{figure}

Meanwhile El Badry et al.\cite{2023MNRAS.518.1057E,2023arXiv230207880E} report on two confirmed sources in the Galactic plane that were first identified by the astrometric motions in Gaia DR3 (see also Chakrabarti et al\cite{2022arXiv221005003C}). Spectroscopic confirmation of radial velocities in these sources identifies them as single-line spectroscopic binaries, and when combined with the astrometric constraints, provides a direct measure of the component masses.\cite{2023MNRAS.518.1057E,2023arXiv230207880E} The first system, Gaia BH1, has a solar-like ($\sim 0.93M_\odot$) main sequence star in a 185.6~d orbit around a $9.62\pm0.18M_\odot$ presumed BH\cite{2023MNRAS.518.1057E}. The second system has an approximately solar-mass giant-branch companion in a 1277~d orbital period about a $8.9\pm 0.3M_\odot$ presumed BH. Just as VFTS 243 is in some ways a longer period analog of the BH-HMXBs, Gaia BH1 and Gaia BH2 are long-period analogs of the BH-LMXBs. Neither Gaia BH1 or BH2 are directly interacting or Roche lobe overflowing at their long orbital periods. As a result, these BHs are quiescent, unlike their shorter-period X-ray counterparts. Figure \ref{fig:MP} shows the BH period--mass distribution and is adapted from El Badry et al.\cite{2023arXiv230207880E}

These first discoveries of BHs in binary systems with wider orbital periods emphasize that the X-ray binary population likely represents a ``tip of the iceberg" in terms of the shortest orbital period binaries containing BHs, thus those that are most likely to drive luminous accretion.

\section{Conclusions and Open Questions}\label{sec:conclusion}

To date, just a handful of the $\sim 10^8$ BHs thought to exist in our Galaxy have been discovered. Many of these have been revealed through signatures of accretion, emitting broadly from Gamma-ray and X-ray to radio wavelengths.\cite{2006ARA&A..44...49R,2010ApJ...725.1918O,2011ApJ...741..103F} More recently, several strong candidates for isolated or non-accreting BHs in binaries have been uncovered through either microlensing or their astrometric motions. The known systems are summarized mass and orbital period space in Figure \ref{fig:MP}, adapted from El Badry et al.\cite{2023arXiv230207880E}

There are a number of ongoing areas of active investigation in the modeling and observation of BHs in the Galaxy. In particular, some key questions include:
\begin{enumerate}
    \item What is the mass function of BH formation? What are the processes that drive it? BHs may form by core collapse of massive stars as well as by compact object mergers  (see Chapter 3 in this volume). These processes combine to shape a BH mass function\cite{2010ApJ...725.1918O}. One important aspect of this mass function is a possible mass gap between the upper NS mass limit and typical BH masses.\cite{2010ApJ...725.1918O,2012ApJ...757...36K,2022arXiv220906844S,2022ApJ...933L..23L,2022AstBu..77..197K} The observed population of BH-HMXBs and BH-LMXBs can be expanded with renewed detection efforts, including more sensitive, broad-band imaging and spectra (at e.g. 3-300 keV), as well as all-sky monitoring (at, e.g. 30 keV to 10 MeV). Also crucial, though quite time-consuming, is the dynamical confirmation of systems through spectroscopic follow up. 

    \item What is the formation history of the BH-LMXBs? And in particular, how do they end up with such low-mass and apparently uniform companion properties (typically $<1M_\odot$ dwarfs)? At the very least, there is tension between the pre-main sequence lifetime of low-mass companions ($10^7$~yr) and the entire evolutionary lifetime of the BH progenitors (few~$\times10^6$~yr). 
   
    \item How do populations of accreting BHs in X-ray emitting systems connect to their detached, more weakly emitting counterparts? Do they represent parts of a continuous distribution or do they come from differing formation channels?  The discovery of a continuum of BH-HMXBs from weak-wind systems like HD 96670\cite{2021ApJ...913...48G} to jet-launching microquasars like SS433\cite{1982Sci...215..247M,1984ARA&A..22..507M,2020A&A...640A..96P} along with a continuum of BHs with low-mass companions from the detached Gaia BH1 and BH2  systems to Roche lobe overflowing BH-LMXBs hints that the BHs we know are part of broad distributions.  
   
    \item How does accretion reshape BH populations? The observations of spins in BH X-ray binaries directly relate to the accreted mass in these systems. How, and why, spins correlate with binary properties like period, mass ratio, and BH mass is a crucial frontier.\cite{2015ApJ...800...17F,2021ARA&A..59..117R} 
    \item What can we learn from the spatial and kinematic distribution of BHs in the galaxy? These properties are already being used to probe natal kicks\cite{2012MNRAS.425.2799R,2017MNRAS.467..298R,2019MNRAS.489.3116A} and correlations between binary properties and kinematics \cite{2019MNRAS.489.3116A}. Our observed populations are also shaped by these spatial distributions\cite{2021ApJ...921..131J}. More sophisticated consideration of these properties may be enabled as a statistical population of BHs grows. 
\end{enumerate}

Outside the Galaxy, or knowledge of BH populations is exploding with the detection of gravitational waves from merging binary BHs. One of the key questions that will drive upcoming research is how populations of gravitational-wave sources and X-ray binaries may relate, as recently posed by Fishbach et al.\cite{2022ApJ...929L..26F} This is also playing out as detached binaries containing BHs and microlensing BHs are now being discovered. Each BH detection technique probes a portion of the full range of BH parameter space. For example, Blackmon and Maccarone\cite{2023MNRAS.519.2995B} have recently shown a relationship between outburst luminosity and period in BH transients, with implications for which binaries we are most likely to discover by monitoring for these outbursts. Jonker et al. caution that extinction in the plane of the Galaxy biases the observed BH-LMXB mass function\cite{2021ApJ...921..131J}. It will be the task of future work to uncover the ways that emergent populations trace the full range of BHs in and beyond our Galaxy.

\section*{Acknowledgements}
We gratefully acknowledge Ariella Katz  for research assistance with the Gaia population of black hole companion counterparts. We thank K. Breivik and K. El-Badry for insight and helpful conversations. 


\begin{thebibliography}{121}
\providecommand{\natexlab}[1]{#1}
\providecommand{\url}[1]{\texttt{#1}}
\expandafter\ifx\csname urlstyle\endcsname\relax
  \providecommand{\doi}[1]{doi: #1}\else
  \providecommand{\doi}{doi: \begingroup \urlstyle{rm}\Url}\fi

\bibitem{1983bhwd.book.....S}
S.~L. {Shapiro} and S.~A. {Teukolsky}, \emph{{Black holes, white dwarfs, and
  neutron stars : the physics of compact objects}}. 1983.

\bibitem{2006csxs.book..507K}
A.~R. {King}.
\newblock {Accretion in compact binaries}.
\newblock In \emph{Compact stellar X-ray sources}, vol.~39, pp. 507--546.
  (2006).

\bibitem{2010csxs.book..157M}
J.~E. {McClintock} and R.~A. {Remillard}.
\newblock {Black hole binaries}.
\newblock In \emph{Compact Stellar X-ray Sources}, p. 157.  (2010).

\bibitem{1965Sci...147..394B}
S.~{Bowyer}, E.~T. {Byram}, T.~A. {Chubb}, and H.~{Friedman}, {Cosmic X-ray
  Sources}, \emph{Science}. {\bf 147}\penalty0 (3656), \penalty0 394--398
  (Jan., 1965).
\newblock \doi{10.1126/science.147.3656.394}.

\bibitem{1971ApJ...165L..27G}
R.~{Giacconi}, E.~{Kellogg}, P.~{Gorenstein}, H.~{Gursky}, and H.~{Tananbaum},
  {An X-Ray Scan of the Galactic Plane from UHURU}, \emph{\apjl}. {\bf 165},
  \penalty0 L27 (Apr., 1971).
\newblock \doi{10.1086/180711}.

\bibitem{1978ApJS...38..357F}
W.~{Forman}, C.~{Jones}, L.~{Cominsky}, P.~{Julien}, S.~{Murray}, G.~{Peters},
  H.~{Tananbaum}, and R.~{Giacconi}, {The fourth Uhuru catalog of X-ray
  sources.}, \emph{\apjs}. {\bf 38}, \penalty0 357--412 (Dec., 1978).
\newblock \doi{10.1086/190561}.

\bibitem{1974ApJ...189L..13R}
R.~E. {Rothschild}, E.~A. {Boldt}, S.~S. {Holt}, and P.~J. {Serlemitsos},
  {Millisecond Temporal Structure in Cygnus X-1}, \emph{\apjl}. {\bf 189},
  \penalty0 L13 (Apr., 1974).
\newblock \doi{10.1086/181452}.

\bibitem{2014wbll.book.....B}
C.~D. {Bailyn}, \emph{{What does a black hole look like?}} 2014.
\newblock \doi{10.1515/9781400850563}.

\bibitem{2006ARA&A..44...49R}
R.~A. {Remillard} and J.~E. {McClintock}, {X-Ray Properties of Black-Hole
  Binaries}, \emph{\araa}. {\bf 44}\penalty0 (1), \penalty0 49--92 (Sept.,
  2006).
\newblock \doi{10.1146/annurev.astro.44.051905.092532}.

\bibitem{1972ApJ...177L...5T}
H.~{Tananbaum}, H.~{Gursky}, E.~{Kellogg}, R.~{Giacconi}, and C.~{Jones},
  {Observation of a Correlated X-Ray Transition in Cygnus X-1}, \emph{\apjl}.
  {\bf 177}, \penalty0 L5 (Oct., 1972).
\newblock \doi{10.1086/181042}.

\bibitem{2018ApJ...860..166T}
R.~E. {Taam}, E.~{Qiao}, B.~F. {Liu}, and E.~{Meyer-Hofmeister}, {A Model for
  Spectral States and Their Transition in Cyg X-1}, \emph{\apj}. 860\penalty0
  (2):\penalty0 166 (June, 2018).
\newblock \doi{10.3847/1538-4357/aac50d}.

\bibitem{1971Natur.233..110M}
P.~{Murdin} and B.~L. {Webster}, {Optical Identification of Cygnus X-1},
  \emph{\nat}. {\bf 233}\penalty0 (5315), \penalty0 110 (Sept., 1971).
\newblock \doi{10.1038/233110a0}.

\bibitem{2011ApJS..193...24S}
A.~{Sota}, J.~{Ma{\'\i}z Apell{\'a}niz}, N.~R. {Walborn}, E.~J. {Alfaro}, R.~H.
  {Barb{\'a}}, N.~I. {Morrell}, R.~C. {Gamen}, and J.~I. {Arias}, {The Galactic
  O-Star Spectroscopic Survey. I. Classification System and Bright Northern
  Stars in the Blue-violet at R \raisebox{-0.5ex}\textasciitilde 2500},
  \emph{\apjs}. 193\penalty0 (2):\penalty0 24 (Apr., 2011).
\newblock \doi{10.1088/0067-0049/193/2/24}.

\bibitem{2018maeb.book.....P}
A.~{Pr{\v{s}}a}, \emph{{Modeling and Analysis of Eclipsing Binary Stars; The
  theory and design principles of PHOEBE}}. 2018.
\newblock \doi{10.1088/978-0-7503-1287-5}.

\bibitem{2021Sci...371.1046M}
J.~C.~A. {Miller-Jones}, A.~{Bahramian}, J.~A. {Orosz}, I.~{Mandel}, L.~{Gou},
  T.~J. {Maccarone}, C.~J. {Neijssel}, X.~{Zhao}, J.~{Zi{\'o}{\l}kowski}, M.~J.
  {Reid}, P.~{Uttley}, X.~{Zheng}, D.-Y. {Byun}, R.~{Dodson}, V.~{Grinberg},
  T.~{Jung}, J.-S. {Kim}, B.~{Marcote}, S.~{Markoff}, M.~J. {Rioja}, A.~P.
  {Rushton}, D.~M. {Russell}, G.~R. {Sivakoff}, A.~J. {Tetarenko}, V.~{Tudose},
  and J.~{Wilms}, {Cygnus X-1 contains a 21-solar mass black
  hole{\textemdash}Implications for massive star winds}, \emph{Science}. {\bf
  371}\penalty0 (6533), \penalty0 1046--1049 (Mar., 2021).
\newblock \doi{10.1126/science.abb3363}.

\bibitem{2011ApJ...742...84O}
J.~A. {Orosz}, J.~E. {McClintock}, J.~P. {Aufdenberg}, R.~A. {Remillard}, M.~J.
  {Reid}, R.~{Narayan}, and L.~{Gou}, {The Mass of the Black Hole in Cygnus
  X-1}, \emph{\apj}. 742\penalty0 (2):\penalty0 84 (Dec., 2011).
\newblock \doi{10.1088/0004-637X/742/2/84}.

\bibitem{2009ApJ...697..573O}
J.~A. {Orosz}, D.~{Steeghs}, J.~E. {McClintock}, M.~A.~P. {Torres},
  I.~{Bochkov}, L.~{Gou}, R.~{Narayan}, M.~{Blaschak}, A.~M. {Levine}, R.~A.
  {Remillard}, C.~D. {Bailyn}, M.~M. {Dwyer}, and M.~{Buxton}, {A New Dynamical
  Model for the Black Hole Binary LMC X-1}, \emph{\apj}. {\bf 697}\penalty0
  (1), \penalty0 573--591 (May, 2009).
\newblock \doi{10.1088/0004-637X/697/1/573}.

\bibitem{2014ApJ...794..154O}
J.~A. {Orosz}, J.~F. {Steiner}, J.~E. {McClintock}, M.~M. {Buxton}, C.~D.
  {Bailyn}, D.~{Steeghs}, A.~{Guberman}, and M.~A.~P. {Torres}, {The Mass of
  the Black Hole in LMC X-3}, \emph{\apj}. 794\penalty0 (2):\penalty0 154
  (Oct., 2014).
\newblock \doi{10.1088/0004-637X/794/2/154}.

\bibitem{2009AJ....138.1243H}
J.~{Harris} and D.~{Zaritsky}, {The Star Formation History of the Large
  Magellanic Cloud}, \emph{\aj}. {\bf 138}\penalty0 (5), \penalty0 1243--1260
  (Nov., 2009).
\newblock \doi{10.1088/0004-6256/138/5/1243}.

\bibitem{2013MNRAS.429L.104Z}
A.~A. {Zdziarski}, J.~{Mikolajewska}, and K.~{Belczynski}, {Cyg X-3: a low-mass
  black hole or a neutron star.}, \emph{\mnras}. {\bf 429}, \penalty0
  L104--L108 (Feb., 2013).
\newblock \doi{10.1093/mnrasl/sls035}.

\bibitem{1984ARA&A..22..507M}
B.~{Margon}, {Observations of SS 433}, \emph{\araa}. {\bf 22}, \penalty0
  507--536 (Jan., 1984).
\newblock \doi{10.1146/annurev.aa.22.090184.002451}.

\bibitem{2018MNRAS.479.4844C}
A.~M. {Cherepashchuk}, K.~A. {Postnov}, and A.~A. {Belinski}, {On masses of the
  components in SS433}, \emph{\mnras}. {\bf 479}\penalty0 (4), \penalty0
  4844--4848 (Oct., 2018).
\newblock \doi{10.1093/mnras/sty1853}.

\bibitem{2020A&A...640A..96P}
P.~{Picchi}, S.~N. {Shore}, E.~J. {Harvey}, and A.~{Berdyugin}, {An optical
  spectroscopic and polarimetric study of the microquasar binary system SS
  433}, \emph{\aap}. 640:\penalty0 A96 (Aug., 2020).
\newblock \doi{10.1051/0004-6361/202037960}.

\bibitem{2021Univ....8...13C}
A.~{Cherepashchuk}, {Progress in Understanding the Nature of SS433},
  \emph{Universe}. {\bf 8}\penalty0 (1), \penalty0 13 (Dec., 2021).
\newblock \doi{10.3390/universe8010013}.

\bibitem{2021ApJ...913...48G}
S.~{Gomez} and J.~E. {Grindlay}, {Optical Analysis and Modeling of HD96670, a
  New Black Hole X-Ray Binary Candidate}, \emph{\apj}. 913\penalty0
  (1):\penalty0 48 (May, 2021).
\newblock \doi{10.3847/1538-4357/abf24c}.

\bibitem{2016A&A...587A..61C}
J.~M. {Corral-Santana}, J.~{Casares}, T.~{Mu{\~n}oz-Darias}, F.~E. {Bauer},
  I.~G. {Mart{\'\i}nez-Pais}, and D.~M. {Russell}, {BlackCAT: A catalogue of
  stellar-mass black holes in X-ray transients}, \emph{\aap}. 587:\penalty0 A61
  (Mar., 2016).
\newblock \doi{10.1051/0004-6361/201527130}.

\bibitem{1975Natur.257..656E}
M.~{Elvis}, C.~G. {Page}, K.~A. {Pounds}, M.~J. {Ricketts}, and M.~J.~L.
  {Turner}, {Discovery of powerful transient X-ray source A0620-00 with Ariel V
  Sky Survey Experiment}, \emph{\nat}. {\bf 257}\penalty0 (5528), \penalty0
  656--657 (Oct., 1975).
\newblock \doi{10.1038/257656a0}.

\bibitem{1986ApJ...308..110M}
J.~E. {McClintock} and R.~A. {Remillard}, {The Black Hole Binary A0620-00},
  \emph{\apj}. {\bf 308}, \penalty0 110 (Sept., 1986).
\newblock \doi{10.1086/164482}.

\bibitem{2016ApJS..222...15T}
B.~E. {Tetarenko}, G.~R. {Sivakoff}, C.~O. {Heinke}, and J.~C. {Gladstone},
  {WATCHDOG: A Comprehensive All-sky Database of Galactic Black Hole X-ray
  Binaries}, \emph{\apjs}. 222\penalty0 (2):\penalty0 15 (Feb., 2016).
\newblock \doi{10.3847/0067-0049/222/2/15}.

\bibitem{1976ApJ...203L..17E}
L.~J. {Eachus}, E.~L. {Wright}, and W.~{Liller}, {Optical observations of the
  recurrent Nova associated with A0620-00 : 1917-1975.}, \emph{\apjl}. {\bf
  203}, \penalty0 L17--L19 (Jan., 1976).
\newblock \doi{10.1086/182009}.

\bibitem{2001NewAR..45..449L}
J.-P. {Lasota}, {The disc instability model of dwarf novae and low-mass X-ray
  binary transients}, \emph{\nar}. {\bf 45}\penalty0 (7), \penalty0 449--508
  (June, 2001).
\newblock \doi{10.1016/S1387-6473(01)00112-9}.

\bibitem{2018MNRAS.474...69A}
K.~{Arur} and T.~J. {Maccarone}, {Selection effects on the orbital period
  distribution of low-mass black hole X-ray binaries}, \emph{\mnras}. {\bf
  474}\penalty0 (1), \penalty0 69--76 (Feb., 2018).
\newblock \doi{10.1093/mnras/stx2762}.

\bibitem{2015ApJ...805...87Y}
Z.~{Yan} and W.~{Yu}, {X-Ray Outbursts of Low-mass X-Ray Binary Transients
  Observed in the RXTE Era}, \emph{\apj}. 805\penalty0 (2):\penalty0 87 (June,
  2015).
\newblock \doi{10.1088/0004-637X/805/2/87}.

\bibitem{2012IAUS..285...29G}
J.~{Grindlay}, S.~{Tang}, E.~{Los}, and M.~{Servillat}.
\newblock {Opening the 100-Year Window for Time-Domain Astronomy}.
\newblock In eds. E.~{Griffin}, R.~{Hanisch}, and R.~{Seaman}, \emph{New
  Horizons in Time Domain Astronomy}, vol. 285, pp. 29--34 (Apr., 2012).
\newblock \doi{10.1017/S1743921312000166}.

\bibitem{2022arXiv220610053B}
A.~{Bahramian} and N.~{Degenaar}, {Low-Mass X-ray Binaries}, \emph{arXiv
  e-prints}. art. arXiv:2206.10053 (June, 2022).
\newblock \doi{10.48550/arXiv.2206.10053}.

\bibitem{2021ApJ...921..131J}
P.~G. {Jonker}, K.~{Kaur}, N.~{Stone}, and M.~A.~P. {Torres}, {The Observed
  Mass Distribution of Galactic Black Hole LMXBs Is Biased against Massive
  Black Holes}, \emph{\apj}. 921\penalty0 (2):\penalty0 131 (Nov., 2021).
\newblock \doi{10.3847/1538-4357/ac2839}.

\bibitem{1999A&A...352L..87N}
G.~{Nelemans}, T.~M. {Tauris}, and E.~P.~J. {van den Heuvel}, {Constraints on
  mass ejection in black hole formation derived from black hole X-ray
  binaries}, \emph{\aap}. {\bf 352}, \penalty0 L87--L90 (Dec., 1999).
\newblock \doi{10.48550/arXiv.astro-ph/9911054}.

\bibitem{2004MNRAS.354..355J}
P.~G. {Jonker} and G.~{Nelemans}, {The distances to Galactic low-mass X-ray
  binaries: consequences for black hole luminosities and kicks}, \emph{\mnras}.
  {\bf 354}\penalty0 (2), \penalty0 355--366 (Oct., 2004).
\newblock \doi{10.1111/j.1365-2966.2004.08193.x}.

\bibitem{2012MNRAS.425.2799R}
S.~{Repetto}, M.~B. {Davies}, and S.~{Sigurdsson}, {Investigating stellar-mass
  black hole kicks}, \emph{\mnras}. {\bf 425}\penalty0 (4), \penalty0
  2799--2809 (Oct., 2012).
\newblock \doi{10.1111/j.1365-2966.2012.21549.x}.

\bibitem{2015MNRAS.453.3341R}
S.~{Repetto} and G.~{Nelemans}, {Constraining the formation of black holes in
  short-period black hole low-mass X-ray binaries}, \emph{\mnras}. {\bf
  453}\penalty0 (3), \penalty0 3341--3355 (Nov., 2015).
\newblock \doi{10.1093/mnras/stv1753}.

\bibitem{2017MNRAS.467..298R}
S.~{Repetto}, A.~P. {Igoshev}, and G.~{Nelemans}, {The Galactic distribution of
  X-ray binaries and its implications for compact object formation and natal
  kicks}, \emph{\mnras}. {\bf 467}\penalty0 (1), \penalty0 298--310 (May,
  2017).
\newblock \doi{10.1093/mnras/stx027}.

\bibitem{2019MNRAS.485.2642G}
P.~{Gandhi}, A.~{Rao}, M.~A.~C. {Johnson}, J.~A. {Paice}, and T.~J.
  {Maccarone}, {Gaia Data Release 2 distances and peculiar velocities for
  Galactic black hole transients}, \emph{\mnras}. {\bf 485}\penalty0 (2),
  \penalty0 2642--2655 (May, 2019).
\newblock \doi{10.1093/mnras/stz438}.

\bibitem{2019MNRAS.489.3116A}
P.~{Atri}, J.~C.~A. {Miller-Jones}, A.~{Bahramian}, R.~M. {Plotkin}, P.~G.
  {Jonker}, G.~{Nelemans}, T.~J. {Maccarone}, G.~R. {Sivakoff}, A.~T. {Deller},
  S.~{Chaty}, M.~A.~P. {Torres}, S.~{Horiuchi}, J.~{McCallum}, T.~{Natusch},
  C.~J. {Phillips}, J.~{Stevens}, and S.~{Weston}, {Potential kick velocity
  distribution of black hole X-ray binaries and implications for natal kicks},
  \emph{\mnras}. {\bf 489}\penalty0 (3), \penalty0 3116--3134 (Nov., 2019).
\newblock \doi{10.1093/mnras/stz2335}.

\bibitem{2020MNRAS.496L..22G}
P.~{Gandhi}, A.~{Rao}, P.~A. {Charles}, K.~{Belczynski}, T.~J. {Maccarone},
  K.~{Arur}, and J.~M. {Corral-Santana}, {A period-dependent spatial scatter of
  Galactic black hole transients}, \emph{\mnras}. {\bf 496}\penalty0 (1),
  \penalty0 L22--L27 (July, 2020).
\newblock \doi{10.1093/mnrasl/slaa081}.

\bibitem{1987A&A...183...47D}
M.~{de Kool}, E.~P.~J. {van den Heuvel}, and E.~{Pylyser}, {An evolutionary
  scenario for the black hole binary A0620-00.}, \emph{\aap}. {\bf 183},
  \penalty0 47--52 (Sept., 1987).

\bibitem{1976IAUS...73...75P}
B.~{Paczynski}.
\newblock {Common Envelope Binaries}.
\newblock In eds. P.~{Eggleton}, S.~{Mitton}, and J.~{Whelan}, \emph{Structure
  and Evolution of Close Binary Systems}, vol.~73, p.~75 (Jan., 1976).

\bibitem{2002ApJ...565.1107P}
P.~{Podsiadlowski}, S.~{Rappaport}, and E.~D. {Pfahl}, {Evolutionary Sequences
  for Low- and Intermediate-Mass X-Ray Binaries}, \emph{\apj}. {\bf
  565}\penalty0 (2), \penalty0 1107--1133 (Feb., 2002).
\newblock \doi{10.1086/324686}.

\bibitem{1993PASP..105.1373I}
J.~{Iben}, Icko and M.~{Livio}, {Common Envelopes in Binary Star Evolution},
  \emph{\pasp}. {\bf 105}, \penalty0 1373 (Dec., 1993).
\newblock \doi{10.1086/133321}.

\bibitem{2013A&ARv..21...59I}
N.~{Ivanova}, S.~{Justham}, X.~{Chen}, O.~{De Marco}, C.~L. {Fryer},
  E.~{Gaburov}, H.~{Ge}, E.~{Glebbeek}, Z.~{Han}, X.~D. {Li}, G.~{Lu},
  T.~{Marsh}, P.~{Podsiadlowski}, A.~{Potter}, N.~{Soker}, R.~{Taam}, T.~M.
  {Tauris}, E.~P.~J. {van den Heuvel}, and R.~F. {Webbink}, {Common envelope
  evolution: where we stand and how we can move forward}, \emph{\aapr}.
  21:\penalty0 59 (Feb., 2013).
\newblock \doi{10.1007/s00159-013-0059-2}.

\bibitem{1994MNRAS.270..121H}
Z.~{Han}, P.~{Podsiadlowski}, and P.~P. {Eggleton}, {A possible criterion for
  envelope ejection in asymptotic giant branch or first giant branch stars.},
  \emph{\mnras}. {\bf 270}, \penalty0 121--130 (Sept., 1994).
\newblock \doi{10.1093/mnras/270.1.121}.

\bibitem{2001A&A...369..170T}
T.~M. {Tauris} and J.~D.~M. {Dewi}, {Research Note On the binding energy
  parameter of common envelope evolution. Dependency on the definition of the
  stellar core boundary during spiral-in}, \emph{\aap}. {\bf 369}, \penalty0
  170--173 (Apr., 2001).
\newblock \doi{10.1051/0004-6361:20010099}.

\bibitem{2016A&A...596A..58K}
M.~U. {Kruckow}, T.~M. {Tauris}, N.~{Langer}, D.~{Sz{\'e}csi}, P.~{Marchant},
  and P.~{Podsiadlowski}, {Common-envelope ejection in massive binary stars.
  Implications for the progenitors of GW150914 and GW151226}, \emph{\aap}.
  596:\penalty0 A58 (Nov., 2016).
\newblock \doi{10.1051/0004-6361/201629420}.

\bibitem{2021A&A...645A..54K}
J.~{Klencki}, G.~{Nelemans}, A.~G. {Istrate}, and M.~{Chruslinska}, {It has to
  be cool: Supergiant progenitors of binary black hole mergers from
  common-envelope evolution}, \emph{\aap}. 645:\penalty0 A54 (Jan., 2021).
\newblock \doi{10.1051/0004-6361/202038707}.

\bibitem{2022MNRAS.511.2326V}
A.~{Vigna-G{\'o}mez}, M.~{Wassink}, J.~{Klencki}, A.~{Istrate}, G.~{Nelemans},
  and I.~{Mandel}, {Stellar response after stripping as a model for
  common-envelope outcomes}, \emph{\mnras}. {\bf 511}\penalty0 (2), \penalty0
  2326--2338 (Apr., 2022).
\newblock \doi{10.1093/mnras/stac237}.

\bibitem{1999ApJ...521..723K}
V.~{Kalogera}, {Donor Stars in Black Hole X-Ray Binaries}, \emph{\apj}. {\bf
  521}\penalty0 (2), \penalty0 723--734 (Aug., 1999).
\newblock \doi{10.1086/307562}.

\bibitem{2016MNRAS.458.4188M}
E.~{Michaely} and H.~B. {Perets}, {Tidal capture formation of low-mass X-ray
  binaries from wide binaries in the field}, \emph{\mnras}. {\bf 458}\penalty0
  (4), \penalty0 4188--4197 (June, 2016).
\newblock \doi{10.1093/mnras/stw368}.

\bibitem{2017MNRAS.469.3088K}
J.~{Klencki}, G.~{Wiktorowicz}, W.~{G{\l}adysz}, and K.~{Belczynski},
  {Dynamical formation of black hole low-mass X-ray binaries in the field: an
  alternative to the common envelope}, \emph{\mnras}. {\bf 469}\penalty0 (3),
  \penalty0 3088--3101 (Aug., 2017).
\newblock \doi{10.1093/mnras/stx842}.

\bibitem{2017ApJS..230...15M}
M.~{Moe} and R.~{Di Stefano}, {Mind Your Ps and Qs: The Interrelation between
  Period (P) and Mass-ratio (Q) Distributions of Binary Stars}, \emph{\apjs}.
  230\penalty0 (2):\penalty0 15 (June, 2017).
\newblock \doi{10.3847/1538-4365/aa6fb6}.

\bibitem{1975MNRAS.172P..15F}
A.~C. {Fabian}, J.~E. {Pringle}, and M.~J. {Rees}, {Tidal capture formation of
  binary systems and X-ray sources in globular clusters.}, \emph{\mnras}. {\bf
  172}, \penalty0 15 (Aug., 1975).
\newblock \doi{10.1093/mnras/172.1.15P}.

\bibitem{1977ApJ...213..183P}
W.~H. {Press} and S.~A. {Teukolsky}, {On formation of close binaries by
  two-body tidal capture.}, \emph{\apj}. {\bf 213}, \penalty0 183--192 (Apr.,
  1977).
\newblock \doi{10.1086/155143}.

\bibitem{2014ApJ...782...60K}
N.~A. {Kaib} and S.~N. {Raymond}, {Very Wide Binary Stars as the Primary Source
  of Stellar Collisions in the Galaxy}, \emph{\apj}. 782\penalty0 (2):\penalty0
  60 (Feb., 2014).
\newblock \doi{10.1088/0004-637X/782/2/60}.

\bibitem{2016ApJ...822L..24N}
S.~{Naoz}, T.~{Fragos}, A.~{Geller}, A.~P. {Stephan}, and F.~A. {Rasio},
  {Formation of Black Hole Low-mass X-Ray Binaries in Hierarchical Triple
  Systems}, \emph{\apjl}. 822\penalty0 (2):\penalty0 L24 (May, 2016).
\newblock \doi{10.3847/2041-8205/822/2/L24}.

\bibitem{2022A&A...661A..61T}
S.~{Toonen}, T.~C.~N. {Boekholt}, and S.~{Portegies Zwart}, {Stellar triples on
  the edge. Comprehensive overview of the evolution of destabilised triples
  leading to stellar and binary exotica}, \emph{\aap}. 661:\penalty0 A61 (May,
  2022).
\newblock \doi{10.1051/0004-6361/202141991}.

\bibitem{2004ApJ...601L.171K}
V.~{Kalogera}, A.~R. {King}, and F.~A. {Rasio}, {Could Black Hole X-Ray
  Binaries Be Detected in Globular Clusters?}, \emph{\apjl}. {\bf 601}\penalty0
  (2), \penalty0 L171--L174 (Feb., 2004).
\newblock \doi{10.1086/382042}.

\bibitem{1987A&A...184..164R}
A.~{Ray}, A.~K. {Kembhavi}, and H.~M. {Antia}, {Evolution of stellar binaries
  formed by tidal capture}, \emph{\aap}. {\bf 184}\penalty0 (1-2), \penalty0
  164--172 (Oct., 1987).

\bibitem{2022ApJ...937...37M}
M.~{MacLeod}, M.~{Vick}, and A.~{Loeb}, {Tidal Wave Breaking in the Eccentric
  Lead-in to Mass Transfer and Common Envelope Phases}, \emph{\apj}.
  937\penalty0 (1):\penalty0 37 (Sept., 2022).
\newblock \doi{10.3847/1538-4357/ac8aff}.

\bibitem{2019MNRAS.484.1506L}
C.~X. {Lu} and S.~{Naoz}, {Supernovae kicks in hierarchical triple systems},
  \emph{\mnras}. {\bf 484}\penalty0 (2), \penalty0 1506--1525 (Apr., 2019).
\newblock \doi{10.1093/mnras/stz036}.

\bibitem{2015ApJ...800...17F}
T.~{Fragos} and J.~E. {McClintock}, {The Origin of Black Hole Spin in Galactic
  Low-mass X-Ray Binaries}, \emph{\apj}. 800\penalty0 (1):\penalty0 17 (Feb.,
  2015).
\newblock \doi{10.1088/0004-637X/800/1/17}.

\bibitem{2011CQGra..28k4009M}
J.~E. {McClintock}, R.~{Narayan}, S.~W. {Davis}, L.~{Gou}, A.~{Kulkarni}, J.~A.
  {Orosz}, R.~F. {Penna}, R.~A. {Remillard}, and J.~F. {Steiner}, {Measuring
  the spins of accreting black holes}, \emph{Classical and Quantum Gravity}.
  28\penalty0 (11):\penalty0 114009 (June, 2011).
\newblock \doi{10.1088/0264-9381/28/11/114009}.

\bibitem{2014SSRv..183..295M}
J.~E. {McClintock}, R.~{Narayan}, and J.~F. {Steiner}, {Black Hole Spin via
  Continuum Fitting and the Role of Spin in Powering Transient Jets},
  \emph{\ssr}. {\bf 183}\penalty0 (1-4), \penalty0 295--322 (Sept., 2014).
\newblock \doi{10.1007/s11214-013-0003-9}.

\bibitem{1989MNRAS.238..729F}
A.~C. {Fabian}, M.~J. {Rees}, L.~{Stella}, and N.~E. {White}, {X-ray
  fluorescence from the inner disc in Cygnus X-1.}, \emph{\mnras}. {\bf 238},
  \penalty0 729--736 (May, 1989).
\newblock \doi{10.1093/mnras/238.3.729}.

\bibitem{2014SSRv..183..277R}
C.~S. {Reynolds}, {Measuring Black Hole Spin Using X-Ray Reflection
  Spectroscopy}, \emph{\ssr}. {\bf 183}\penalty0 (1-4), \penalty0 277--294
  (Sept., 2014).
\newblock \doi{10.1007/s11214-013-0006-6}.

\bibitem{2021ARA&A..59..117R}
C.~S. {Reynolds}, {Observational Constraints on Black Hole Spin}, \emph{\araa}.
  {\bf 59}, \penalty0 117--154 (Sept., 2021).
\newblock \doi{10.1146/annurev-astro-112420-035022}.

\bibitem{1986ApJ...304....1P}
B.~{Paczynski}, {Gravitational Microlensing by the Galactic Halo}, \emph{\apj}.
  {\bf 304}, \penalty0 1 (May, 1986).
\newblock \doi{10.1086/164140}.

\bibitem{1991ApJ...371L..63P}
B.~{Paczynski}, {Gravitational Microlensing of the Galactic Bulge Stars},
  \emph{\apjl}. {\bf 371}, \penalty0 L63 (Apr., 1991).
\newblock \doi{10.1086/186003}.

\bibitem{1993ASPC...43..291A}
C.~{Alcock}, R.~A. {Allsman}, T.~S. {Axelrod}, D.~P. {Bennett}, K.~H. {Cook},
  H.~S. {Park}, S.~L. {Marshall}, C.~W. {Stubbs}, K.~{Griest}, S.~{Perlmutter},
  W.~{Sutherland}, K.~C. {Freeman}, B.~A. {Peterson}, P.~J. {Quinn}, and A.~W.
  {Rodgers}.
\newblock {The MACHO Project - a Search for the Dark Matter in the Milky-Way}.
\newblock In ed. B.~T. {Soifer}, \emph{Sky Surveys. Protostars to
  Protogalaxies}, vol.~43, \emph{Astronomical Society of the Pacific Conference
  Series}, p. 291 (Jan., 1993).

\bibitem{2019ApJS..244...29M}
P.~{Mr{\'o}z}, A.~{Udalski}, J.~{Skowron}, M.~K. {Szyma{\'n}ski},
  I.~{Soszy{\'n}ski}, {\L}.~{Wyrzykowski}, P.~{Pietrukowicz},
  S.~{Koz{\l}owski}, R.~{Poleski}, K.~{Ulaczyk}, K.~{Rybicki}, and P.~{Iwanek},
  {Microlensing Optical Depth and Event Rate toward the Galactic Bulge from 8
  yr of OGLE-IV Observations}, \emph{\apjs}. 244\penalty0 (2):\penalty0 29
  (Oct., 2019).
\newblock \doi{10.3847/1538-4365/ab426b}.

\bibitem{1993Natur.365..621A}
C.~{Alcock}, C.~W. {Akerlof}, R.~A. {Allsman}, T.~S. {Axelrod}, D.~P.
  {Bennett}, S.~{Chan}, K.~H. {Cook}, K.~C. {Freeman}, K.~{Griest}, S.~L.
  {Marshall}, H.~S. {Park}, S.~{Perlmutter}, B.~A. {Peterson}, M.~R. {Pratt},
  P.~J. {Quinn}, A.~W. {Rodgers}, C.~W. {Stubbs}, and W.~{Sutherland},
  {Possible gravitational microlensing of a star in the Large Magellanic
  Cloud}, \emph{\nat}. {\bf 365}\penalty0 (6447), \penalty0 621--623 (Oct.,
  1993).
\newblock \doi{10.1038/365621a0}.

\bibitem{2000ApJ...542..281A}
C.~{Alcock}, R.~A. {Allsman}, D.~R. {Alves}, T.~S. {Axelrod}, A.~C. {Becker},
  D.~P. {Bennett}, K.~H. {Cook}, N.~{Dalal}, A.~J. {Drake}, K.~C. {Freeman},
  M.~{Geha}, K.~{Griest}, M.~J. {Lehner}, S.~L. {Marshall}, D.~{Minniti}, C.~A.
  {Nelson}, B.~A. {Peterson}, P.~{Popowski}, M.~R. {Pratt}, P.~J. {Quinn},
  C.~W. {Stubbs}, W.~{Sutherland}, A.~B. {Tomaney}, T.~{Vandehei}, and
  D.~{Welch}, {The MACHO Project: Microlensing Results from 5.7 Years of Large
  Magellanic Cloud Observations}, \emph{\apj}. {\bf 542}\penalty0 (1),
  \penalty0 281--307 (Oct., 2000).
\newblock \doi{10.1086/309512}.

\bibitem{2002ApJ...576L.131A}
E.~{Agol}, M.~{Kamionkowski}, L.~V.~E. {Koopmans}, and R.~D. {Blandford},
  {Finding Black Holes with Microlensing}, \emph{\apjl}. {\bf 576}\penalty0
  (2), \penalty0 L131--L135 (Sept., 2002).
\newblock \doi{10.1086/343758}.

\bibitem{2002ApJ...579..639B}
D.~P. {Bennett}, A.~C. {Becker}, J.~L. {Quinn}, A.~B. {Tomaney}, C.~{Alcock},
  R.~A. {Allsman}, D.~R. {Alves}, T.~S. {Axelrod}, J.~J. {Calitz}, K.~H.
  {Cook}, A.~J. {Drake}, P.~C. {Fragile}, K.~C. {Freeman}, M.~{Geha},
  K.~{Griest}, B.~R. {Johnson}, S.~C. {Keller}, C.~{Laws}, M.~J. {Lehner},
  S.~L. {Marshall}, D.~{Minniti}, C.~A. {Nelson}, B.~A. {Peterson},
  P.~{Popowski}, M.~R. {Pratt}, P.~J. {Quinn}, S.~H. {Rhie}, C.~W. {Stubbs},
  W.~{Sutherland}, T.~{Vandehei}, D.~{Welch}, {MACHO Collaboration}, and {MPS
  Collaboration}, {Gravitational Microlensing Events Due to Stellar-Mass Black
  Holes}, \emph{\apj}. {\bf 579}\penalty0 (2), \penalty0 639--659 (Nov., 2002).
\newblock \doi{10.1086/342225}.

\bibitem{2016MNRAS.458.3012W}
{\L}.~{Wyrzykowski}, Z.~{Kostrzewa-Rutkowska}, J.~{Skowron}, K.~A. {Rybicki},
  P.~{Mr{\'o}z}, S.~{Koz{\l}owski}, A.~{Udalski}, M.~K. {Szyma{\'n}ski},
  G.~{Pietrzy{\'n}ski}, I.~{Soszy{\'n}ski}, K.~{Ulaczyk}, P.~{Pietrukowicz},
  R.~{Poleski}, M.~{Pawlak}, K.~{I{\l}kiewicz}, and N.~J. {Rattenbury}, {Black
  hole, neutron star and white dwarf candidates from microlensing with
  OGLE-III}, \emph{\mnras}. {\bf 458}\penalty0 (3), \penalty0 3012--3026 (May,
  2016).
\newblock \doi{10.1093/mnras/stw426}.

\bibitem{2022ApJ...933L..23L}
C.~Y. {Lam}, J.~R. {Lu}, A.~{Udalski}, I.~{Bond}, D.~P. {Bennett},
  J.~{Skowron}, P.~{Mr{\'o}z}, R.~{Poleski}, T.~{Sumi}, M.~K. {Szyma{\'n}ski},
  S.~{Koz{\l}owski}, P.~{Pietrukowicz}, I.~{Soszy{\'n}ski}, K.~{Ulaczyk},
  {\L}.~{Wyrzykowski}, S.~{Miyazaki}, D.~{Suzuki}, N.~{Koshimoto}, N.~J.
  {Rattenbury}, M.~W. {Hosek}, F.~{Abe}, R.~{Barry}, A.~{Bhattacharya},
  A.~{Fukui}, H.~{Fujii}, Y.~{Hirao}, Y.~{Itow}, R.~{Kirikawa}, I.~{Kondo},
  Y.~{Matsubara}, S.~{Matsumoto}, Y.~{Muraki}, G.~{Olmschenk}, C.~{Ranc},
  A.~{Okamura}, Y.~{Satoh}, S.~I. {Silva}, T.~{Toda}, P.~J. {Tristram},
  A.~{Vandorou}, H.~{Yama}, N.~S. {Abrams}, S.~{Agarwal}, S.~{Rose}, and S.~K.
  {Terry}, {An Isolated Mass-gap Black Hole or Neutron Star Detected with
  Astrometric Microlensing}, \emph{\apjl}. 933\penalty0 (1):\penalty0 L23
  (July, 2022).
\newblock \doi{10.3847/2041-8213/ac7442}.

\bibitem{2000ApJ...535..928G}
A.~{Gould}, {Measuring the Remnant Mass Function of the Galactic Bulge},
  \emph{\apj}. {\bf 535}\penalty0 (2), \penalty0 928--931 (June, 2000).
\newblock \doi{10.1086/308865}.

\bibitem{2022ApJ...930..159A}
J.~J. {Andrews} and V.~{Kalogera}, {Constraining Black Hole Natal Kicks with
  Astrometric Microlensing}, \emph{\apj}. 930\penalty0 (2):\penalty0 159 (May,
  2022).
\newblock \doi{10.3847/1538-4357/ac66d6}.

\bibitem{2022ApJ...933...83S}
K.~C. {Sahu}, J.~{Anderson}, S.~{Casertano}, H.~E. {Bond}, A.~{Udalski},
  M.~{Dominik}, A.~{Calamida}, A.~{Bellini}, T.~M. {Brown}, M.~{Rejkuba},
  V.~{Bajaj}, N.~{Kains}, H.~C. {Ferguson}, C.~L. {Fryer}, P.~{Yock},
  P.~{Mr{\'o}z}, S.~{Koz{\l}owski}, P.~{Pietrukowicz}, R.~{Poleski},
  J.~{Skowron}, I.~{Soszy{\'n}ski}, M.~K. {Szyma{\'n}ski}, K.~{Ulaczyk},
  {\L}.~{Wyrzykowski}, R.~K. {Barry}, D.~P. {Bennett}, I.~A. {Bond},
  Y.~{Hirao}, S.~I. {Silva}, I.~{Kondo}, N.~{Koshimoto}, C.~{Ranc}, N.~J.
  {Rattenbury}, T.~{Sumi}, D.~{Suzuki}, P.~J. {Tristram}, A.~{Vandorou}, J.-P.
  {Beaulieu}, J.-B. {Marquette}, A.~{Cole}, P.~{Fouqu{\'e}}, K.~{Hill},
  S.~{Dieters}, C.~{Coutures}, D.~{Dominis-Prester}, C.~{Bennett},
  E.~{Bachelet}, J.~{Menzies}, M.~{Albrow}, K.~{Pollard}, A.~{Gould}, J.~C.
  {Yee}, W.~{Allen}, L.~A. {Almeida}, G.~{Christie}, J.~{Drummond},
  A.~{Gal-Yam}, E.~{Gorbikov}, F.~{Jablonski}, C.-U. {Lee}, D.~{Maoz},
  I.~{Manulis}, J.~{McCormick}, T.~{Natusch}, R.~W. {Pogge}, Y.~{Shvartzvald},
  U.~G. {J{\o}rgensen}, K.~A. {Alsubai}, M.~I. {Andersen}, V.~{Bozza}, S.~C.
  {Novati}, M.~{Burgdorf}, T.~C. {Hinse}, M.~{Hundertmark}, T.-O. {Husser},
  E.~{Kerins}, P.~{Longa-Pe{\~n}a}, L.~{Mancini}, M.~{Penny}, S.~{Rahvar},
  D.~{Ricci}, S.~{Sajadian}, J.~{Skottfelt}, C.~{Snodgrass}, J.~{Southworth},
  J.~{Tregloan-Reed}, J.~{Wambsganss}, O.~{Wertz}, Y.~{Tsapras}, R.~A.
  {Street}, D.~M. {Bramich}, K.~{Horne}, I.~A. {Steele}, and {RoboNet
  Collaboration}, {An Isolated Stellar-mass Black Hole Detected through
  Astrometric Microlensing}, \emph{\apj}. 933\penalty0 (1):\penalty0 83 (July,
  2022).
\newblock \doi{10.3847/1538-4357/ac739e}.

\bibitem{2022ApJS..260...55L}
C.~Y. {Lam}, J.~R. {Lu}, A.~{Udalski}, I.~{Bond}, D.~P. {Bennett},
  J.~{Skowron}, P.~{Mr{\'o}z}, R.~{Poleski}, T.~{Sumi}, M.~K. {Szyma{\'n}ski},
  S.~{Koz{\l}owski}, P.~{Pietrukowicz}, I.~{Soszy{\'n}ski}, K.~{Ulaczyk},
  {\L}.~{Wyrzykowski}, S.~{Miyazaki}, D.~{Suzuki}, N.~{Koshimoto}, N.~J.
  {Rattenbury}, M.~W. {Hosek}, F.~{Abe}, R.~{Barry}, A.~{Bhattacharya},
  A.~{Fukui}, H.~{Fujii}, Y.~{Hirao}, Y.~{Itow}, R.~{Kirikawa}, I.~{Kondo},
  Y.~{Matsubara}, S.~{Matsumoto}, Y.~{Muraki}, G.~{Olmschenk}, C.~{Ranc},
  A.~{Okamura}, Y.~{Satoh}, S.~I. {Silva}, T.~{Toda}, P.~J. {Tristram},
  A.~{Vandorou}, H.~{Yama}, N.~S. {Abrams}, S.~{Agarwal}, S.~{Rose}, and S.~K.
  {Terry}, {Supplement: ``An Isolated Mass-gap Black Hole or Neutron Star
  Detected with Astrometric Microlensing'' (2022, ApJL, 933, L23)},
  \emph{\apjs}. 260\penalty0 (2):\penalty0 55 (June, 2022).
\newblock \doi{10.3847/1538-4365/ac7441}.

\bibitem{2022ApJ...937L..24M}
P.~{Mr{\'o}z}, A.~{Udalski}, and A.~{Gould}, {Systematic Errors as a Source of
  Mass Discrepancy in Black Hole Microlensing Event OGLE-2011-BLG-0462},
  \emph{\apjl}. 937\penalty0 (2):\penalty0 L24 (Oct., 2022).
\newblock \doi{10.3847/2041-8213/ac90bb}.

\bibitem{2022A&A...664A.159M}
L.~{Mahy}, H.~{Sana}, T.~{Shenar}, K.~{Sen}, N.~{Langer}, P.~{Marchant},
  M.~{Abdul-Masih}, G.~{Banyard}, J.~{Bodensteiner}, D.~M. {Bowman},
  K.~{Dsilva}, M.~{Fabry}, C.~{Hawcroft}, S.~{Janssens}, T.~{Van Reeth}, and
  C.~{Eldridge}, {Identifying quiescent compact objects in massive Galactic
  single-lined spectroscopic binaries}, \emph{\aap}. 664:\penalty0 A159 (Aug.,
  2022).
\newblock \doi{10.1051/0004-6361/202243147}.

\bibitem{2022NatAs...6.1085S}
T.~{Shenar}, H.~{Sana}, L.~{Mahy}, K.~{El-Badry}, P.~{Marchant}, N.~{Langer},
  C.~{Hawcroft}, M.~{Fabry}, K.~{Sen}, L.~A. {Almeida}, M.~{Abdul-Masih},
  J.~{Bodensteiner}, P.~A. {Crowther}, M.~{Gieles}, M.~{Gromadzki},
  V.~{H{\'e}nault-Brunet}, A.~{Herrero}, A.~{de Koter}, P.~{Iwanek},
  S.~{Koz{\l}owski}, D.~J. {Lennon}, J.~{Ma{\'\i}z Apell{\'a}niz},
  P.~{Mr{\'o}z}, A.~F.~J. {Moffat}, A.~{Picco}, P.~{Pietrukowicz},
  R.~{Poleski}, K.~{Rybicki}, F.~R.~N. {Schneider}, D.~M. {Skowron},
  J.~{Skowron}, I.~{Soszy{\'n}ski}, M.~K. {Szyma{\'n}ski}, S.~{Toonen},
  A.~{Udalski}, K.~{Ulaczyk}, J.~S. {Vink}, and M.~{Wrona}, {An X-ray-quiet
  black hole born with a negligible kick in a massive binary within the Large
  Magellanic Cloud}, \emph{Nature Astronomy}. {\bf 6}, \penalty0 1085--1092
  (July, 2022).
\newblock \doi{10.1038/s41550-022-01730-y}.

\bibitem{2018MNRAS.475L..15G}
B.~{Giesers}, S.~{Dreizler}, T.-O. {Husser}, S.~{Kamann}, G.~{Anglada
  Escud{\'e}}, J.~{Brinchmann}, C.~M. {Carollo}, M.~M. {Roth}, P.~M.
  {Weilbacher}, and L.~{Wisotzki}, {A detached stellar-mass black hole
  candidate in the globular cluster NGC 3201}, \emph{\mnras}. {\bf
  475}\penalty0 (1), \penalty0 L15--L19 (Mar., 2018).
\newblock \doi{10.1093/mnrasl/slx203}.

\bibitem{2020A&A...635A..65K}
S.~{Kamann}, B.~{Giesers}, N.~{Bastian}, J.~{Brinchmann}, S.~{Dreizler},
  F.~{G{\"o}ttgens}, T.~O. {Husser}, M.~{Latour}, P.~M. {Weilbacher}, and
  L.~{Wisotzki}, {The binary content of multiple populations in NGC 3201},
  \emph{\aap}. 635:\penalty0 A65 (Mar., 2020).
\newblock \doi{10.1051/0004-6361/201936843}.

\bibitem{2023MNRAS.518.1057E}
K.~{El-Badry}, H.-W. {Rix}, E.~{Quataert}, A.~W. {Howard}, H.~{Isaacson},
  J.~{Fuller}, K.~{Hawkins}, K.~{Breivik}, K.~W.~K. {Wong}, A.~C. {Rodriguez},
  C.~{Conroy}, S.~{Shahaf}, T.~{Mazeh}, F.~{Arenou}, K.~B. {Burdge},
  D.~{Bashi}, S.~{Faigler}, D.~R. {Weisz}, R.~{Seeburger}, S.~{Almada Monter},
  and J.~{Wojno}, {A Sun-like star orbiting a black hole}, \emph{\mnras}. {\bf
  518}\penalty0 (1), \penalty0 1057--1085 (Jan., 2023).
\newblock \doi{10.1093/mnras/stac3140}.

\bibitem{2023arXiv230207880E}
K.~{El-Badry}, H.-W. {Rix}, Y.~{Cendes}, A.~C. {Rodriguez}, C.~{Conroy},
  E.~{Quataert}, K.~{Hawkins}, E.~{Zari}, M.~{Hobson}, K.~{Breivik}, A.~{Rau},
  E.~{Berger}, S.~{Shahaf}, R.~{Seeburger}, K.~B. {Burdge}, D.~W. {Latham},
  L.~A. {Buchhave}, A.~{Bieryla}, D.~{Bashi}, T.~{Mazeh}, and S.~{Faigler}, {A
  red giant orbiting a black hole}, \emph{arXiv e-prints}. art.
  arXiv:2302.07880 (Feb., 2023).
\newblock \doi{10.48550/arXiv.2302.07880}.

\bibitem{2022arXiv221005003C}
S.~{Chakrabarti}, J.~D. {Simon}, P.~A. {Craig}, H.~{Reggiani},
  P.~{Guhathakurta}, P.~A. {Dalba}, E.~N. {Kirby}, P.~{Chang}, D.~R. {Hey},
  A.~{Savino}, and M.~{Geha}, {A non-interacting Galactic black hole candidate
  in a binary system with a main-sequence star}, \emph{arXiv e-prints}. art.
  arXiv:2210.05003 (Oct., 2022).
\newblock \doi{10.48550/arXiv.2210.05003}.

\bibitem{2000MNRAS.317..528W}
N.~A. {Webb}, T.~{Naylor}, Z.~{Ioannou}, P.~A. {Charles}, and T.~{Shahbaz}, {A
  TiO study of the black hole binary GRO J0422+32 in a very low state},
  \emph{\mnras}. {\bf 317}\penalty0 (3), \penalty0 528--534 (Sept., 2000).
\newblock \doi{10.1046/j.1365-8711.2000.03608.x}.

\bibitem{2010ApJ...710.1127C}
A.~G. {Cantrell}, C.~D. {Bailyn}, J.~A. {Orosz}, J.~E. {McClintock}, R.~A.
  {Remillard}, C.~S. {Froning}, J.~{Neilsen}, D.~M. {Gelino}, and L.~{Gou},
  {The Inclination of the Soft X-Ray Transient A0620-00 and the Mass of its
  Black Hole}, \emph{\apj}. {\bf 710}\penalty0 (2), \penalty0 1127--1141 (Feb.,
  2010).
\newblock \doi{10.1088/0004-637X/710/2/1127}.

\bibitem{2003ApJ...599.1254G}
D.~M. {Gelino} and T.~E. {Harrison}, {GRO J0422+32: The Lowest Mass Black
  Hole?}, \emph{\apj}. {\bf 599}\penalty0 (2), \penalty0 1254--1259 (Dec.,
  2003).
\newblock \doi{10.1086/379311}.

\bibitem{2013AJ....145...21K}
J.~{Khargharia}, C.~S. {Froning}, E.~L. {Robinson}, and D.~M. {Gelino}, {The
  Mass of the Black Hole in XTE J1118+480}, \emph{\aj}. 145\penalty0
  (1):\penalty0 21 (Jan., 2013).
\newblock \doi{10.1088/0004-6256/145/1/21}.

\bibitem{2016ApJ...825...46W}
J.~{Wu}, J.~A. {Orosz}, J.~E. {McClintock}, I.~{Hasan}, C.~D. {Bailyn},
  L.~{Gou}, and Z.~{Chen}, {The Mass of the Black Hole in the X-ray Binary Nova
  Muscae 1991}, \emph{\apj}. 825\penalty0 (1):\penalty0 46 (July, 2016).
\newblock \doi{10.3847/0004-637X/825/1/46}.

\bibitem{2021MNRAS.506..581M}
D.~{Mata S{\'a}nchez}, A.~{Rau}, A.~{{\'A}lvarez Hern{\'a}ndez}, T.~F.~J. {van
  Grunsven}, M.~A.~P. {Torres}, and P.~G. {Jonker}, {Dynamical confirmation of
  a stellar mass black hole in the transient X-ray dipping binary MAXI
  J1305-704}, \emph{\mnras}. {\bf 506}\penalty0 (1), \penalty0 581--594 (Sept.,
  2021).
\newblock \doi{10.1093/mnras/stab1714}.

\bibitem{2011ApJ...730...75O}
J.~A. {Orosz}, J.~F. {Steiner}, J.~E. {McClintock}, M.~A.~P. {Torres}, R.~A.
  {Remillard}, C.~D. {Bailyn}, and J.~M. {Miller}, {An Improved Dynamical Model
  for the Microquasar XTE J1550-564}, \emph{\apj}. 730\penalty0 (2):\penalty0
  75 (Apr., 2011).
\newblock \doi{10.1088/0004-637X/730/2/75}.

\bibitem{2002ApJ...568..845O}
J.~A. {Orosz}, P.~J. {Groot}, M.~{van der Klis}, J.~E. {McClintock}, M.~R.
  {Garcia}, P.~{Zhao}, R.~K. {Jain}, C.~D. {Bailyn}, and R.~A. {Remillard},
  {Dynamical Evidence for a Black Hole in the Microquasar XTE J1550-564},
  \emph{\apj}. {\bf 568}\penalty0 (2), \penalty0 845--861 (Apr., 2002).
\newblock \doi{10.1086/338984}.

\bibitem{2004ApJ...616..376O}
J.~A. {Orosz}, J.~E. {McClintock}, R.~A. {Remillard}, and S.~{Corbel}, {Orbital
  Parameters for the Black Hole Binary XTE J1650-500}, \emph{\apj}. {\bf
  616}\penalty0 (1), \penalty0 376--382 (Nov., 2004).
\newblock \doi{10.1086/424892}.

\bibitem{2003MNRAS.339.1031S}
T.~{Shahbaz}, {Determining the spectroscopic mass ratio in interacting
  binaries: application to X-Ray Nova Sco 1994}, \emph{\mnras}. {\bf
  339}\penalty0 (4), \penalty0 1031--1040 (Mar., 2003).
\newblock \doi{10.1046/j.1365-8711.2003.06258.x}.

\bibitem{1997AJ....114.1170H}
E.~T. {Harlaftis}, D.~{Steeghs}, K.~{Horne}, and A.~V. {Filippenko}, {A doppler
  map and mass-ration constraint for the black-hole x-ray nova ophiuchi 1977.},
  \emph{\aj}. {\bf 114}, \penalty0 1170--1175 (Sept., 1997).
\newblock \doi{10.1086/118548}.

\bibitem{2017ApJ...846..132H}
M.~{Heida}, P.~G. {Jonker}, M.~A.~P. {Torres}, and A.~{Chiavassa}, {The Mass
  Function of GX 339-4 from Spectroscopic Observations of Its Donor Star},
  \emph{\apj}. 846\penalty0 (2):\penalty0 132 (Sept., 2017).
\newblock \doi{10.3847/1538-4357/aa85df}.

\bibitem{2020ApJ...893L..37T}
M.~A.~P. {Torres}, J.~{Casares}, F.~{Jim{\'e}nez-Ibarra},
  A.~{{\'A}lvarez-Hern{\'a}ndez}, T.~{Mu{\~n}oz-Darias}, M.~{Armas Padilla},
  P.~G. {Jonker}, and M.~{Heida}, {The Binary Mass Ratio in the Black Hole
  Transient MAXI J1820+070}, \emph{\apjl}. 893\penalty0 (2):\penalty0 L37
  (Apr., 2020).
\newblock \doi{10.3847/2041-8213/ab863a}.

\bibitem{2022MNRAS.517.1476Y}
I.~V. {Yanes-Rizo}, M.~A.~P. {Torres}, J.~{Casares}, S.~E. {Motta},
  T.~{Mu{\~n}oz-Darias}, P.~{Rodr{\'\i}guez-Gil}, M.~{Armas Padilla},
  F.~{Jim{\'e}nez-Ibarra}, P.~G. {Jonker}, J.~M. {Corral-Santana}, and
  R.~{Fender}, {A refined dynamical mass for the black hole in the X-ray
  transient XTE J1859+226}, \emph{\mnras}. {\bf 517}\penalty0 (1), \penalty0
  1476--1482 (Nov., 2022).
\newblock \doi{10.1093/mnras/stac2719}.

\bibitem{2014ApJ...784....2M}
R.~K.~D. {MacDonald}, C.~D. {Bailyn}, M.~{Buxton}, A.~G. {Cantrell},
  R.~{Chatterjee}, R.~{Kennedy-Shaffer}, J.~A. {Orosz}, C.~B. {Markwardt}, and
  J.~H. {Swank}, {The Black Hole Binary V4641 Sagitarii: Activity in Quiescence
  and Improved Mass Determinations}, \emph{\apj}. 784\penalty0 (1):\penalty0 2
  (Mar., 2014).
\newblock \doi{10.1088/0004-637X/784/1/2}.

\bibitem{2014ApJ...796....2R}
M.~J. {Reid}, J.~E. {McClintock}, J.~F. {Steiner}, D.~{Steeghs}, R.~A.
  {Remillard}, V.~{Dhawan}, and R.~{Narayan}, {A Parallax Distance to the
  Microquasar GRS 1915+105 and a Revised Estimate of its Black Hole Mass},
  \emph{\apj}. 796\penalty0 (1):\penalty0 2 (Nov., 2014).
\newblock \doi{10.1088/0004-637X/796/1/2}.

\bibitem{2010ApJ...716.1105K}
J.~{Khargharia}, C.~S. {Froning}, and E.~L. {Robinson}, {Near-infrared
  Spectroscopy of Low-mass X-ray Binaries: Accretion Disk Contamination and
  Compact Object Mass Determination in V404 Cyg and Cen X-4}, \emph{\apj}. {\bf
  716}\penalty0 (2), \penalty0 1105--1117 (June, 2010).
\newblock \doi{10.1088/0004-637X/716/2/1105}.

\bibitem{2004AJ....127..481I}
Z.~{Ioannou}, E.~L. {Robinson}, W.~F. {Welsh}, and C.~A. {Haswell}, {The Mass
  of the Black Hole in GS 2000+25}, \emph{\aj}. {\bf 127}\penalty0 (1),
  \penalty0 481--488 (Jan., 2004).
\newblock \doi{10.1086/380215}.

\bibitem{2010ApJ...725.1918O}
F.~{{\"O}zel}, D.~{Psaltis}, R.~{Narayan}, and J.~E. {McClintock}, {The Black
  Hole Mass Distribution in the Galaxy}, \emph{\apj}. {\bf 725}\penalty0 (2),
  \penalty0 1918--1927 (Dec., 2010).
\newblock \doi{10.1088/0004-637X/725/2/1918}.

\bibitem{2011ApJ...741..103F}
W.~M. {Farr}, N.~{Sravan}, A.~{Cantrell}, L.~{Kreidberg}, C.~D. {Bailyn},
  I.~{Mandel}, and V.~{Kalogera}, {The Mass Distribution of Stellar-mass Black
  Holes}, \emph{\apj}. 741\penalty0 (2):\penalty0 103 (Nov., 2011).
\newblock \doi{10.1088/0004-637X/741/2/103}.

\bibitem{2012ApJ...757...36K}
L.~{Kreidberg}, C.~D. {Bailyn}, W.~M. {Farr}, and V.~{Kalogera}, {Mass
  Measurements of Black Holes in X-Ray Transients: Is There a Mass Gap?},
  \emph{\apj}. 757\penalty0 (1):\penalty0 36 (Sept., 2012).
\newblock \doi{10.1088/0004-637X/757/1/36}.

\bibitem{2022arXiv220906844S}
J.~C. {Siegel}, I.~{Kiato}, V.~{Kalogera}, C.~P.~L. {Berry}, T.~J. {Maccarone},
  K.~{Breivik}, J.~J. {Andrews}, S.~S. {Bavera}, A.~{Dotter}, T.~{Fragos},
  K.~{Kovlakas}, D.~{Misra}, K.~A. {Rocha}, P.~M. {Srivastava}, M.~{Sun},
  Z.~{Xing}, and E.~{Zapartas}, {Investigating the Lower Mass Gap with Low Mass
  X-ray Binary Population Synthesis}, \emph{arXiv e-prints}. art.
  arXiv:2209.06844 (Sept., 2022).
\newblock \doi{10.48550/arXiv.2209.06844}.

\bibitem{2022AstBu..77..197K}
N.~{Kumar} and V.~V. {Sokolov}, {Mass Distribution and ``Mass Gap'' of Compact
  Stellar Remnants in Binary Systems}, \emph{Astrophysical Bulletin}. {\bf
  77}\penalty0 (2), \penalty0 197--213 (June, 2022).
\newblock \doi{10.1134/S1990341322020043}.

\bibitem{1982Sci...215..247M}
B.~{Margon}, {Relativistic Jets in SS 433}, \emph{Science}. {\bf 215}\penalty0
  (4530), \penalty0 247--252 (Jan., 1982).
\newblock \doi{10.1126/science.215.4530.247}.

\bibitem{2022ApJ...929L..26F}
M.~{Fishbach} and V.~{Kalogera}, {Apples and Oranges: Comparing Black Holes in
  X-Ray Binaries and Gravitational-wave Sources}, \emph{\apjl}. 929\penalty0
  (2):\penalty0 L26 (Apr., 2022).
\newblock \doi{10.3847/2041-8213/ac64a5}.

\bibitem{2023MNRAS.519.2995B}
V.~A. {Blackmon} and T.~J. {Maccarone}, {Relating peak optical luminosity and
  orbital period of stellar-mass black holes in X-ray binaries}, \emph{\mnras}.
  {\bf 519}\penalty0 (2), \penalty0 2995--2999 (Feb., 2023).
\newblock \doi{10.1093/mnras/stac3680}.

\end{thebibliography}
\bibliographystyle{ws-rv-van}

\end{document}